\documentclass[aps,prl,twocolumn,showpacs,superscriptaddress,floatfix]{revtex4-2}
\usepackage[T1]{fontenc}
\usepackage{graphicx}
\usepackage{dcolumn}
\usepackage{bm}
\usepackage{float}
\usepackage{amssymb}
\usepackage{amsmath,amssymb}
\usepackage{dsfont}
\usepackage{hyperref}
\usepackage{xcolor}
\hypersetup{colorlinks=true,citecolor={blue},linkcolor={blue},urlcolor={blue}}
\usepackage{soul}
\usepackage{upgreek}

\usepackage{mathrsfs}
\usepackage{lmodern}
\usepackage[utf8]{inputenc}
\usepackage{natbib}
\usepackage{hyperref}
\usepackage{color}
\usepackage{units}
\usepackage[english]{babel}
\usepackage{svg}

\setcitestyle{super}
\usepackage[normalem]{ulem}
\usepackage{titlesec}
\titleformat{\section}[display]{\bf\large}{}{0pt}{}
\titleformat{\subsection}[display]{\bf}{}{0pt}{}
\titlespacing\section{0pt}{12pt plus 4pt minus 2pt}{0pt plus 2pt minus 2pt}
\titlespacing\subsection{0pt}{12pt plus 4pt minus 2pt}{0pt plus 2pt minus 2pt}
\titlespacing\subsubsection{0pt}{12pt plus 4pt minus 2pt}{0pt plus 2pt minus 2pt}

\begin{document}

\title{Antiferromagnetic Ising model in a triangular vortex-lattice of quantum fluids of light}

\author{Sergey Alyatkin}\email{s.alyatkin@skoltech.ru}
\affiliation{Hybrid Photonics Laboratory, Skolkovo Institute of Science and Technology, Territory of Innovation Center Skolkovo, Bolshoy Boulevard 30, building 1, 121205 Moscow, Russia}

\author{Carles Milián}
\affiliation{Institut Universitari de Matem\`{a}tica Pura i Aplicada, Universitat Polit\`{e}cnica de Val\`{e}ncia, 46022 Val\`{e}ncia, Spain}

\author{Yaroslav V. Kartashov}
\affiliation{Institute of Spectroscopy of Russian Academy of Sciences, Fizicheskaya Str., 5, Troitsk, Moscow, 108840, Russia}

\author{Kirill A. Sitnik}
\affiliation{Hybrid Photonics Laboratory, Skolkovo Institute of Science and Technology, Territory of Innovation Center Skolkovo, Bolshoy Boulevard 30, building 1, 121205 Moscow, Russia}

\author{Ivan Gnusov}
\affiliation{Hybrid Photonics Laboratory, Skolkovo Institute of Science and Technology, Territory of Innovation Center Skolkovo, Bolshoy Boulevard 30, building 1, 121205 Moscow, Russia}

\author{Julian D.  T\"opfer}
\affiliation{Hybrid Photonics Laboratory, Skolkovo Institute of Science and Technology, Territory of Innovation Center Skolkovo, Bolshoy Boulevard 30, building 1, 121205 Moscow, Russia}

\author{Helgi Sigur{\dh}sson}
\affiliation{Institute of Experimental Physics, Faculty of Physics, University of Warsaw, ul. Pasteura 5, PL-02-093 Warsaw, Poland}
\affiliation{Science Institute, University of Iceland, Dunhagi 3, IS-107, Reykjavik, Iceland}

\author{Pavlos G. Lagoudakis}\email{pavlos.lagoudakis@gmail.com}
\affiliation{Hybrid Photonics Laboratory, Skolkovo Institute of Science and Technology, Territory of Innovation Center Skolkovo, Bolshoy Boulevard 30, building 1, 121205 Moscow, Russia}

\date{\today}

\begin{abstract} 
Vortices are topologically distinctive objects appearing as phase twists in coherent fields of optical beams and Bose-Einstein condensates. Structured networks and artificial lattices of coupled vortices could offer a powerful platform to study and simulate key interaction mechanisms between constituents of condensed matter systems, such as antiferromagnetic interactions, by replacement of spin-angular-momentum with orbital-angular-momentum. Here, we realize such a platform using a macroscopic quantum fluid of light based on microcavity exciton-polariton condensates. We imprint all-optical hexagonal lattice that results into a triangular vortex lattice with each cell possessing a vortex state of charge $l = \pm 1$. We reveal that pairs of coupled condensates spontaneously arrange their orbital-angular-momentum antiparallel, implying a form of artificial orbital ``antiferromagnetism''. We discover that correlation exists between the emergent vortex patterns in large-scale triangular condensate lattices and the low-energy solutions of the corresponding antiferromagnetic Ising system. Our study offers a path toward spontaneously ordered vortex arrays with nearly arbitrary configurations and controlled couplings.
\end{abstract}

\maketitle

\section{Introduction}
Arrays of quantized vortices, appearing as phase-singular twists in wavefunctions, exemplify fascinating self-organising phenomena, originally studied in macroscopic interacting Bose gases such as rotating superfluids and Bose-Einstein condensates displaying nucleation and ordering of single-charge vortices~\cite{Shaeer_Science2001, Fetter_RMP2009}. More recently, optical vortex arrays in lasers have also gained increased interest~\cite{Brambilla1991, Scheuer1999, Chu2012, Piccardo_NatPhot2022} due to the promising applications of optical vortices~\cite{Desyatnikov2005} and phase-singular optics~\cite{DENNIS2009293, Shen_LMAppl2019} in particle manipulation, high resolution imaging, and quantum communication protocols~\cite{Qiu_Science2017, Ni_Science2021}. While optical systems offer superior spatial programmability over said vortex arrays~\cite{Piccardo_NatPhot2022} they do not possess the large interaction strengths inherent to quantum fluids, which play a crucial role in phase transitions and pattern formation~\cite{Zhao_PRB2017}. For this reason, numerous studies have been focused on hybrid systems, which combine the best properties of light and matter to explore vortices~\cite{Rosen_RevModPhys2022}. Here, we investigate such a light-matter platform in the strong-coupling regime based on cavity exciton-polariton (hereafter {\it polariton}) condensates~\cite{Carusotto_RMP2013} offering a compromise between optical programmability~\cite{Topfer_Optica} and intrinsic evolution of vortex arrays in driven-dissipative quantum fluids~\cite{Keeling_PRL2008, Tosi2012, Gao_PRL2018, Ma2020, Sitnik_PRL}. 

Polariton condensates~\cite{Carusotto_RMP2013} are bosonic quantum fluids, characterised by large interaction strengths and light effective mass, that can be all-optically driven into artificial lattices using structured pump patterns~\cite{Pickup_NatComm2020, Alyatkin_NatComm, Pieczarka_Optica2021}. A notable feature of polariton condensates is their nonequilibrium superfluid character~\cite{Amo_NatPhys2009, Sanvitto_NatPhys2010, Lerario_NatPhys2017}, and ability to form vortices that are pinned by sample disorder~\cite{Lagoudakis2008} or through optical trapping~\cite{Dall_PRL2014, Gao_PRLchiral2018, Ma2020}, or spun-up~\cite{Gnusov_2023}. To date, polariton vortex arrays have been observed in the interference pattern of multiple condensate modes~\cite{Tosi2012, Sitnik_PRL} or in specially patterned cavities~\cite{Gao_PRL2018}. However, large-scale programmable vortex lattices with tunable coupling strengths, where each vortex site can be individually addressed, remain unexplored in polariton fluids.

\begin{figure}[t]
\includegraphics[width=0.98\linewidth]{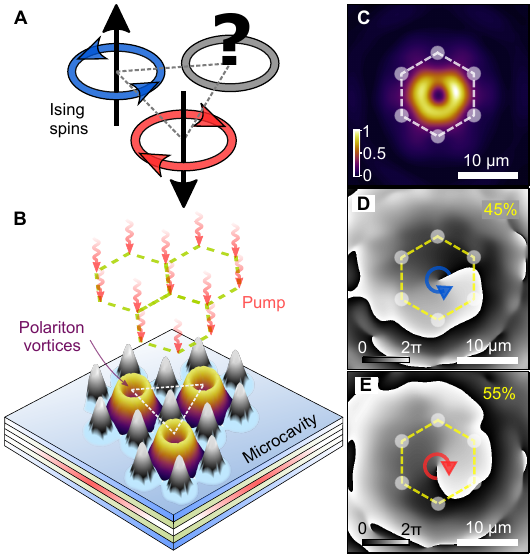}
\caption{\label{fig:epsart}\textbf{Driven trapped polariton condensates with OAM $l = \pm 1$ resembling classical spins}. (\textbf{A}) Example of frustrated triangle for the Ising spins with AFM coupling. (\textbf{B}) Schematic of three polariton condensates in vortex states (yellow-violet surfaces), localized in the potential minima of the pump-induced energy landscape (black-grey surface).  (\textbf{C}) Normalized experimentally measured time-averaged real-space polariton condensate photoluminescence for a single cell. (\textbf{D}),(\textbf{E}) Examples of corresponding phase maps, revealing vortex and antivortex states, respectively. Dashed lines with circles schematically denote the excitation pattern. Numbers, in yellow, indicate statistical occurrence of states over one hundred single-shot realizations.}
\label{fig1}
\end{figure}

As a particular example, we explore here the concept of driven-dissipative geometric frustration by designing a triangular lattice of "antiferromagnetically" (AFM) coupled polariton vortices. Here, AFM refers to polariton orbital-angular-momentum (OAM) instead of spin-angular-momentum. Conventionally, geometric frustration can be described as an inability of a system to minimize its real energy through reduction of pairwise interactions between constituent elements~\cite{vedmedenko2007competing}. A prominent example are Ising spins arranged into an AFM triangular graph~\cite{Wannier_PR1950}, as shown schematically by the black arrows in Fig.~\ref{fig1}A. Only two pairwise interactions can be minimized at any time leaving the third at a higher energy. So far, artificial spin ice systems are popular candidates to explore frustration at large scale~\cite{Skjaervo_NatRevPhys2020} but recently lattices in bosonic systems have steered towards this direction using ultracold atoms~\cite{Struck_science}, coupled lasers~\cite{Nixon_PRL2013} and exciton-polariton condensates~\cite{Ohadi_PRX, Berloff_2017, Cookson2021,Tao_2022}. However, instead of minimizing their real energy, polariton condensates minimize their losses (i.e. maximize their imaginary energy component) when they ballistically couple over mutually pumped region~\cite{Berloff_2017,Cookson2021,Tao_2022}. From this perspective, one can define frustration for polaritons as their inability to minimize losses through non-Hermitian interactions between lattice nodes~\cite{Brunner_NanoPho2020}. 

In this article, we report on the evidence of frustration in a driven-dissipative triangular lattice of polariton condensates. To realize this, we optically imprint a lattice of trapped polariton condensates that populate a vortex state at each lattice node, as shown schematically in Fig.~\ref{fig1}B. Instead of spin-angular-momentum, we work with condensates carrying OAM, with topological charges $l = \pm 1$, defined by a superposition of the traps first excited dipole modes $| P_x \rangle \pm i | P_y \rangle$. We show that we can tune the coupling between synchronized vortex pairs to favour either parallel or antiparallel OAM. The latter constitutes an effective AFM coupling mechanism which results in frustration when triangular geometry of condensates is introduced, due to incommensurate symmetries between the $\mathcal{C}_6$ (sixfold rotational symmetry of the triangular lattice) and $\mathrm{SO}(4)$ (rotation of the two-component OAM condensate) operators. Scaling up to a large 22-vortex system we observe spontaneous formation of non-periodic vortex configurations above threshold, changing from realization to realization, implying existence of multiple available modes. We provide experimental evidence that observed vortex patterns correlate significantly with the ground state solutions of the corresponding Ising system. The large nonlinearities inherent to exciton-polariton systems are crucial for the stability of the vortices, facilitating the readout of their OAM configurations even over long excitation pulses ($\approx10^7$ longer than the polariton lifetime).
\begin{figure*}
\includegraphics[width=0.98\linewidth]{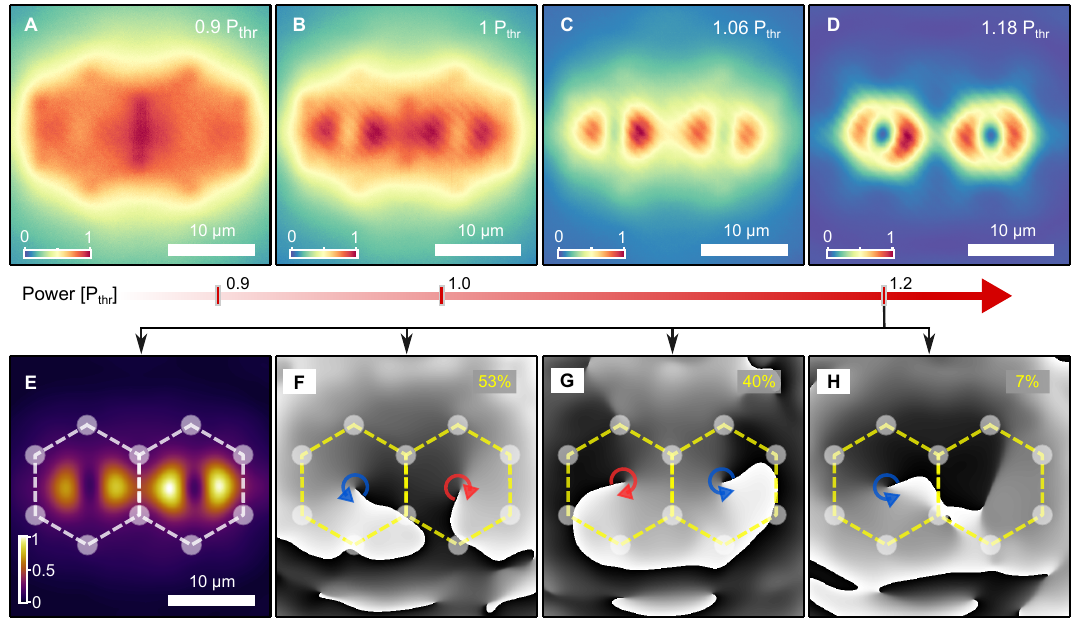}
\caption{\label{fig:epsart}\textbf{Vorticity onset and formation of vortex-antivortex pair in 2-cell structure.} (\textbf{A})-(\textbf{D}) Normalized experimentally measured time-averaged real-space polariton PL as a function of pump power. (\textbf{A}) Below and (\textbf{B}) at threshold the polariton PL is mostly coming from the ridge of the pump-induced traps with no clear vortex formation. (\textbf{C}) At higher power the condensate collapses inside the traps forming first dipole-dipole state, and then (\textbf{D}) a vortex-antivortex state. (\textbf{E}) Measured real-space polariton PL at $P=1.2P_\mathrm{thr}$ and (\textbf{F})-(\textbf{H}) corresponding examples of phase maps. Dashed lines with circles schematically denote the excitation pattern. Numbers, in yellow, in (\textbf{F})-(\textbf{H}) indicate statistical occurrence of states over one hundred single-shot realizations.}
\label{fig2}
\end{figure*}

\section{Results}
We use a planar 2$\lambda$ GaAs-based microcavity with embedded InGaAs quantum wells~\cite{Cilibrizzi_APL2014} held at $\approx$4 K in a closed-cycle helium cryostat. Using a reflective phase-only spatial light modulator (SLM), we transform an incident nonresonant single-mode continuous-wave laser beam into an ordered honeycomb array (the dual of the triangular lattice) of small Gaussian beams focused onto the microcavity plane (see schematic in Fig.~\ref{fig1}B)~\cite{Topfer_Optica}. The nonresonant circularly polarized pump is tuned at the first Bragg minimum of the microcavity reflectivity stop-band to avoid heating of the sample (1.5578 eV). As we show in the Supplementary Material, at the pump power used in experiments, trapped polariton condensate inherits the polarization state from the pump even under nonresonant excitation~\cite{Gnusov2020}. This allows us to excite a single energy state corresponding to polaritons with a predominant pseudo-spin and make therefore, a ``clean'' system without energy splitting (between $\upsigma^+$ and $\upsigma^-$ components). 

The short distance between the pump spots ($d=6.6~\upmu \textrm{m}$) is chosen so as to obtain trapped polariton condensates in the centre of each cell (i.e. in the region between six pumping spots)~\cite{Alyatkin_NatComm}. Physically, polaritons become trapped in this region due to their strong repulsive interactions with the pump-injected background exciton reservoir~\cite{Askitopoulos2013, Askitopoulos_PRB2015, Schneider_2016}. In Supplementary Note 1 we show that for larger honeycomb cells ($d=8.8~\upmu \textrm{m}$) the polaritons mostly condense on top of the pump spots instead of becoming trapped, underlining the importance of tuning the pump parameters. 

\subsection{Vorticity onset in a single trap}
In order to construct spatially extended triangular structure with coupled vortex states, we first implement and analyze their formation in a single, a pair, and a triangle of cells. In particular, the single cell experiment allows us to confirm that the formation of vorticity is truly a spontaneous symmetry-breaking event with a random sign of vortex charge from realization-to-realization. Figure~\ref{fig1}C shows time-integrated real-space polariton photoluminescence (PL) with a donut shape intensity profile for a single cell pumped at $P=1.2P_\mathrm{thr}$. 

The formation of a macroscopic wavefunction $\Psi$ implies irrotational flow everywhere except at the vortex core, where the phase singularity is located and particle density vanishes. Phase winding around the core is quantized in units of $2 \pi l$, where $l \in \mathbb{Z}$. In order to verify this, we apply a homodyne interferometric technique~\cite{Alyatkin2020} to extract the phase of the condensate wavefunction under single-shot excitation conditions (i.e. the sample is excited by a single pulse of 50 $\upmu$s duration).
Our measurements of the single-shot phase maps across different realizations, where vortices form, reveal that 45\% of the shots resulted in formation of a vortex ($l = +1$, Fig.~\ref{fig1}D), and 55\% in an antivortex ($l=-1$, Fig.~\ref{fig1}E). Due to finite negligible pump anisotropy and sample disorder, we also find that 18 realizations out of 100 result in distributions with step-like $\pi$ phase jumps, indicating the formation of a dipolar state, rather than of single vortex state, as it occurs in other 82 realizations, as described in Supplementary Note 2. The absence of a predominant sign of the observed vortex charges means that imprinted excitation pattern does not break chiral symmetry and that vorticity emerges spontaneously during the polariton condensation process. 

In the Materials and Methods section we show, using a simple variational generalised Gross-Pitaevskii model, that $l=\pm1$ vortices in an ideal cylindrically symmetric optical trap (approximating the actual trap induced by six pump spots) are the only stable condensate solutions, in qualitative agreement with our observations. We emphasise that our nonresonant excitation pattern (shown in fig.~S3) does not carry any OAM, nor does it imprint any phase onto the condensate, in contrast to resonant excitation schemes~\cite{Boulier_PRL2016}. 

\subsection{Two coupled vortices}
Next, we focus on two neighbouring cells, shown in Fig.~\ref{fig2}, which convincingly demonstrate signatures of coupling and synchronization. Figures~\ref{fig2}A-D show measured time-averaged real-space polariton PL as a function of pump power. Clearly below (Fig.~\ref{fig2}A) and at condensation threshold (Fig.~\ref{fig2}B) the polariton PL does not reveal univocal signatures of vortex formation. However, at slightly higher power ($P=1.06P_\mathrm{thr}$) the condensate collapses inside the traps forming dipoles (Fig.~\ref{fig2}C), oriented head-to-tails (with parallel nodal lines) in a $\sigma$-bonding fashion as predicted recently~\cite{Cherotchenko_PRB2021}. Then, above some vortex-threshold power the condensates in both cells reveal a donut-shaped intensity distributions (Fig.~\ref{fig2}D), indicating the onset of vorticity. 

In order to verify that the coupling in this configuration is effectively AFM, we extract the phase maps of the vortex pairs at $P=1.2P_\mathrm{thr}$ (see PL in Fig.~\ref{fig2}E). Figures~\ref{fig2}F-H show three typical examples of corresponding phase maps observed in a hundred independent single-shots. Remarkably, we obtained that 93\% of the condensate realizations formed with opposite OAM between the cells. The remaining 7\% correspond to vortex-dipole (or antivortex-dipole) states. Therefore, our data analysis conclusively proves that the vortex coupling between cells strongly favours AFM order. Physically, this can be understood as an optimal constructive interference condition between the traps which maximizes the gain of the condensate pair~\cite{Cherotchenko_PRB2021}. 

The advantage of our optical technique is that by changing some parameters of the pump profile we can force the system to occupy vortex-vortex state (i.e., OAM "ferromagnetic" coupling) instead of vortex-antivortex states. To implement this, we optically imprinted weak potential barriers in the middle of the 2-cell structure in the same spirit as in demonstrated earlier coupling control approach~\cite{Alyatkin2020}. As a result, we managed to demonstrate reversible switching the system from vortex-antivortex to vortex-vortex state as described in Supplementary Note 4.

\begin{figure}
\includegraphics[width=0.98\linewidth]{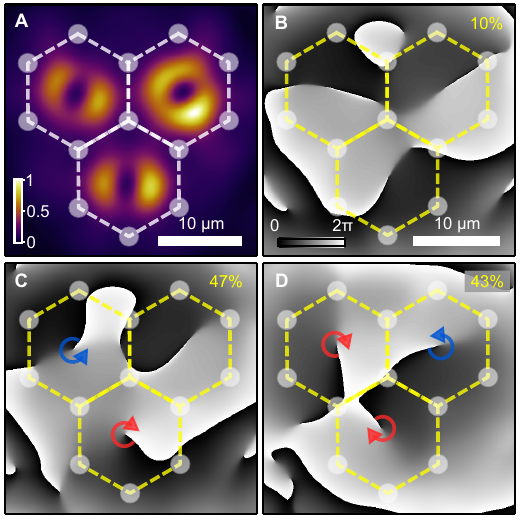}
\caption{\label{fig:epsart} 
\textbf{Vortex states in 3-cell structure.} (\textbf{A}) Normalized experimentally measured time-averaged real-space polariton PL. (\textbf{B})-(\textbf{D}) Corresponding extracted phase maps with numbers, in yellow, indicating statistical occurrence of states over one hundred single-shot realizations. Dashed lines with circles schematically denote the excitation pattern.}
\label{fig3}
\end{figure}

\subsection{Three vortices in a triangle}
Figure~\ref{fig3}A shows the cavity PL after we optically construct an equilateral triangle of trapped condensates. The system's preference for the AFM ordering as well as signatures of the expected topological vortex charge frustration is observed through homodyne interferometric measurements. Details on the interferometric technique are given in Supplementary Note 3 with an example of interference pattern given in fig.~S5. In Fig.~\ref{fig3}B-D we provide few examples of extracted phase maps corresponding to different configurations even though the excitation conditions (geometry and pump power) are kept the same. For example, in Fig.~\ref{fig3}B one can see $\pi$ phase jump lines implying formation of dipole states inside the cells in this particular realization. Dipole states can appear because of the sensitivity of polaritons to inherent cavity disorder or slight pump inhomogeneities which pin the dipoles along a specific direction~\cite{Askitopoulos_PRB2015}. We note that when a condensate is in a perfect superposition of vortex and antivortex modes then the resulting dipole state has zero net OAM and therefore, cannot be assigned to discrete variable of +1 or -1 following the Ising model. However, our further analysis indicates that when there is finite OAM in individual condensate sites, projecting it on classical binary variables shows pattern formation reminiscent of a frustrated triangular Ising system.

None-the-less, for a major part of individual system's realizations (47\% out of one hundred realizations), we find vortex-antivortex (antivortex-vortex) pair accompanied by a dipole state in the third cell, as shown in Fig.~\ref{fig3}C. In 43\% of condensate realizations we reveal vortex-antivortex-vortex (or antivortex-vortex-antivortex) states in all cells, as depicted in Fig.~\ref{fig3}D. We stress that we never observed vortices with the same OAM at the same time in all three cells. We also stress that Fig.~\ref{fig3}C,D show examples over multiple possible arrangements of these type of patterns. These results, in the same vein as those of two cell structure, clearly indicate the tendency for development of the AFM configuration in each pair of neighbouring cells. The fact that multiple realizations display a cell with an indeterminate vortex (see top right cell in Fig.~\ref{fig3}C) implies dynamic indecisiveness reminiscent of frustration. 
\begin{figure}
\includegraphics[width=0.98\linewidth]{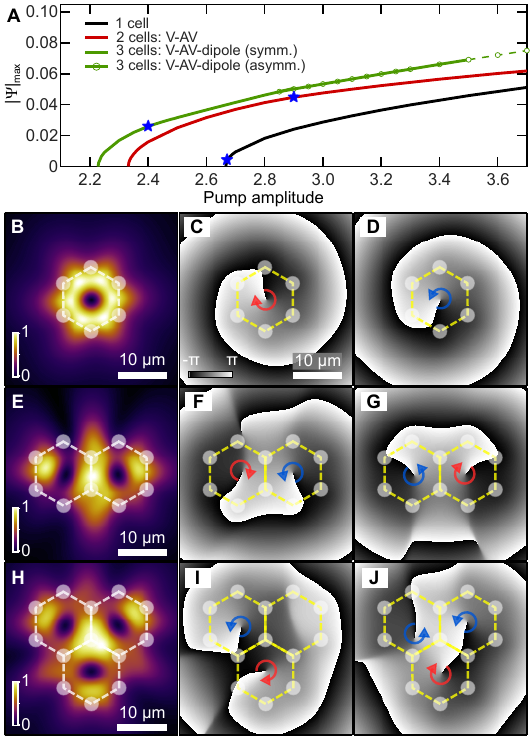}
\caption{\label{fig:epsart} 
\textbf{Vorticity dependence on pump power.} (\textbf{A}) Calculated dependence of the stationary amplitude of the polariton wavefunction emerging in one, two, and three cells with varying pump power. The threshold (start of solid curves) lowers when more pump spots are introduced. The blue stars denote the onset of vorticity with increasing power. Below these stars the polaritons dominantly localize on top of the pump spots with no apparent vorticity. For the 3-cell triangle (green curve) an additional asymmetric frustrated state appears (green circles) at higher powers. (\textbf{B})-(\textbf{J}) Example solutions from 2DGPE simulation of the condensate density and phase in building blocks of the lattice showing agreement with experiment.}
\label{fig4}
\end{figure}

\subsection{Mean field modelling}
Our experimental observations can be reproduced through numerical mean-field simulations, based on the generalised two-dimensional Gross-Pitaevskii equation (2DGPE) describing a macroscopic polariton wavefunction, $\Psi(\textbf{r},t)$, coupled to the rate equation for the exciton reservoir, $n_X(\textbf{r},t)$ (see Materials and Methods). First, we analyze the vorticity onset in the cells of the lattice. In order to obtain families of possible solutions (modes) corresponding to one-, two- and three-cell pump configurations, we apply the Newton method and perform linear stability analysis. Families of stationary states $\Psi(\textbf{r},t)=w(\textbf{r})e^{-i\varepsilon t}$ with $\partial_t n_{X}=0$ (here $\varepsilon$ is the energy detuning) that emerge as stable attractors around threshold correspond to the coloured curves in Fig.~\ref{fig4}A. In this graph we show the dependence of the scaled peak amplitude of the wavefunction $(g_cm_\textrm{eff} r_0^2/\hbar^2)^{1/2}|\Psi|$ on pump power $P_0$. Here, $r_0=1~\upmu \textrm{m}$ is the characteristic spatial scale, $m_\mathrm{eff}$ is the polariton mass, and $g_c$ is the polariton-polariton interaction strength. One can see that the condensation threshold power (starting point of the curves) decreases with increasing number of pumped cells as expected~\cite{Cristofolini_PRL2013}. 

On each solid curve, the onset of vorticity with pump power (i.e., finite net clockwise or anticlockwise rotation) is indicated with a blue star. For a single honeycomb pump cell (black curve), a stable vortex emerges almost at condensation threshold in good agreement with experiments. For the two-cell configuration (red curve), we find that the stable branch emerging from threshold corresponds indeed to vortex-antivortex pairs forming at a critical power given by the blue star, whilst below this power vorticity is absent and field distribution exhibits dipole structure. This critical pump power represents a vorticity threshold, observed experimentally between Fig.~\ref{fig2}C and \ref{fig2}D. When we add a third pump cell, the first stationary family emanating from threshold is associated with a symmetric vortex-antivortex-dipole configuration appearing at the blue star on the green curve in Fig.~\ref{fig4}A, in agreement with observations in Fig.~\ref{fig3}A,C. For lower powers (below blue star) the condensate resembles a triple dipole state, whereas for high powers it gradually transforms into an asymmetric vortex-antivortex-dipole solution (solid green curve with dots). Even further increase of pump power results in unstable behaviour (denoted with green dashed curve in Fig.~\ref{fig4}A) and the emergence of different stationary states $w(\textbf{r})$, such as the antivortex-vortex-antivortex shown in Fig.~\ref{fig3}D. Corresponding solutions around the blue stars on these three branches are shown in Fig.~\ref{fig4}B-J obtained from direct numerical integration of the 2DGPE using white noise initial conditions and damped boundary conditions.

In the Materials and Methods section we additionally develop a simplified model describing the coupling between vortices by treating them as localised orbitals in each trap (i.e., tight binding approach). This reduces the problem of modelling a 2+1 nonlinear partial differential equation into a set of $2N$ ordinary differential equations~\eqref{eq.cpm}. Here $N$ is the number of traps and the factor "2" appears because of clockwise and anticlockwise currents. This effective "spinor" model allows us to analytically prove the stability of vortices in isolated traps through construction of a Lyapunov function. It also correctly predicts increased power of the vortex threshold when two traps are coupled together, in agreement with experiment and 2DGPE simulations. One advantage of the model~\eqref{2osc} is that it possesses analogies with conventional spinor polaritons in patterned cavities with effective spin-orbit-coupling (subject to future work).

\subsection{Triangular lattice of AFM coupled vortices: evidence of low energy Ising order}

\begin{figure}
\includegraphics[width=0.98\linewidth]{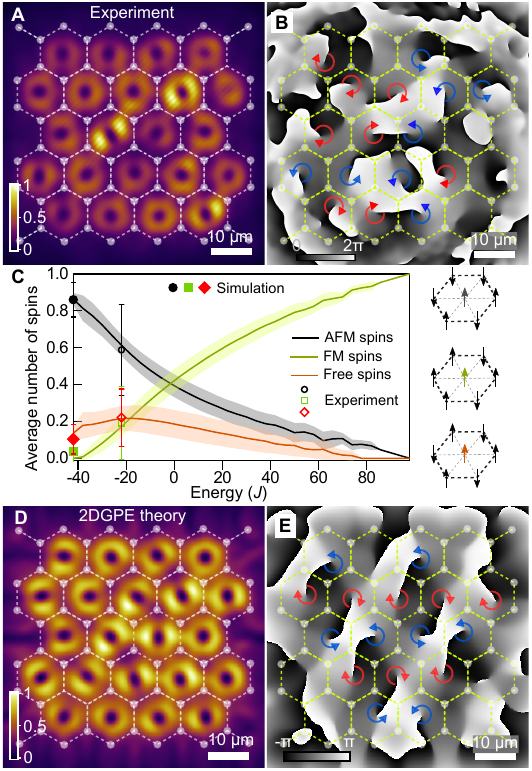}
\caption{\label{fig:epsart} 
\textbf{Vorticity dependence on pump power.} (\textbf{A}) Experimental time-integrated real-space polariton PL intensity (above condensation threshold) and (\textbf{B}) measured single-shot realization of the condensate phase illustrating the formation of a large-scale vortex lattice. (\textbf{C}) Calculated normalized mean number of AFM, FM, and free spins in the 22-spin AFM triangular Ising lattice as a function of energy. The shaded areas correspond to standard deviation (SD). The three types of spins are schematically depicted on the right. The average vortex populations obtained from experiment is plotted with the empty markers with error bars denoting SD. The obtained average populations from a discretized Gross-Pitaevskii simulations are shown with filled black, green and red markers and, remarkably, correspond to the minimum energy $E=-42 J$, see Supplementary Note 6 for details on model and calculations. Overlaid honeycomb lattice in (\textbf{A}),(\textbf{B}),(\textbf{D}),(\textbf{E}) schematically shown with dashed lines and semi-transparent circles indicates the pump positions. Blue and red arrows show schematically the phase winding direction in each cell. (\textbf{D}) Time-averaged realization of the condensate density above threshold reveals donut-shaped PL. (\textbf{E}) Instantaneous condensate phase obtained from mean field modelling confirms vortices formation.}

\label{fig5}
\end{figure}

Finally, we study the vorticity patterns appearing in much larger spatially extended structures,  consisting of 22 cells in a triangular geometry. Figure~\ref{fig5}A shows polariton condensation into a macroscopic coherent state containing multiple vortices across the lattice for a pump power above condensation threshold unlike previous observations in square~\cite{Ohadi_2018Synchr,Berloff_2017,Topfer_Optica}, triangular~\cite{Topfer_Optica}, and Lieb~\cite{Alyatkin_NatComm} optical lattices. Similar to our observations in smaller structures, each cell possesses a donut shaped PL intensity profile. 

Figure~\ref{fig5}B shows extracted from experiment a single-shot phase map with multiple single-charge ($l = \pm 1$) phase singularities across the lattice, schematically marked by blue and red arrows. We underline that the excitation pulse width (50 $\upmu$s) is much longer than the cavity lifetime ($\approx$ 5.5 ps), indicating the topological robustness of the vortices. We observe that across multiple realizations (provided in Supplementary Note 5) the pattern of vortices changes as one might expect due to the spontaneous symmetry breaking upon condensation. 

The apparently random configurations of vortices from realization-to-realization would suggest that the system is completely stochastic, which would be the case if the traps are very weakly coupled. However, as we know from observations on the two-cell (Fig.~\ref{fig2}) and three-cell (Fig.~\ref{fig3}) experiments, the traps are AFM coupled which should manifest in some form of AFM order in the lattice OAM with a highly degenerate "ground state" in analogy to Ising systems~\cite{Wannier_PR1950}.

To address this question, we analyzed the spatial OAM order of the vortices appearing in 25 single-shot realizations of the 22-cell structure, and checked for correlations with the low energy configurations of the Ising Hamiltonian, see Supplementary Note 8 for derivation of its role in picking optimal vortex arrangements,
\begin{equation} \label{eq.Ising}
E = -J \sum_{\langle n ,m \rangle} \sigma_n \sigma_m
\end{equation}
where $J<0$ is the AFM coupling energy, the variables $\sigma_n \in \{\pm1\}$ represent Ising spins and the sum is taken over nearest neighbours in the triangular lattice. The up/down OAM of each lattice node is then projected onto its corresponding Ising spin and classified into one of three groups~\cite{Wannier_PR1950}: (1) spin has more co-aligned neighbours ("FM" spins, $E$ is raised); (2) spin has more counter-aligned neighbours ("AFM" spins, $E$ is lowered); (3) spin has equal co- and counter-aligned neighbours ("free" spins, $E$ is unchanged). Here, we refer the reader to small insets in Fig.~\ref{fig5}C, on the right. The ground state of~\eqref{eq.Ising} can be organised in many different ways using conditionally ordered AFM spins and free spins. 

All $2^{22}$ possible spin configurations and associated energies from equation~\eqref{eq.Ising} can be easily found by brute force. Binning each spin of each configuration into one of the three possible categories described above, we can plot the mean and standard deviation (SD) of the normalized number of spins in each category as a function of Ising energy $E$ (see curves and shaded areas in Fig.~\ref{fig5}C). As expected, the highest energy state has all spins parallel (only FM spins) with zero SD whereas the ground state has only AFM and free spins with finite SD (implying degeneracy). We then apply the same classification to the vortices in our experimental data (see fig.~S8), and extract corresponding statistical occurrence of FM, AFM, and free vortices normalised against the number of detected vortices and plot on top of Fig.~\ref{fig5}C. The three data points (as black, green and red empty markers) are placed where their error is minimum from the solid curves which puts them at $E=-22J$. Remarkably, our data points are within the lower $\approx$6\% of Ising spin configurations from the ground state. From this we conclude, that our vortex lattice does not display stochastic arrangement of vortices, which would give $E=0$, but instead tends towards AFM order.         

Figures~\ref{fig5}D and~\ref{fig5}E show a representative example of the simulated condensate density and phase distributions emerging from random initial conditions, evidently illustrating the formation of a vortex lattice. Excitation from random noise typically yields dynamical, breathing, structures that nevertheless are characterised by persistence of vortices, once they form in the cells. In agreement with the experiment, the arrangement of vortices changes from simulation to simulation (see more examples in fig.~S9), illustrating the highly multi-modal nature of the lattice, yet maintaining a dominant AFM order across the 22-cell structure.

\section{Discussion}
We note that self-arranging vortices are well-known in rotating superfluids~\cite{donnelly1991quantized}, atomic BECs~\cite{Fetter_RMP2009}, and in nonlinear optical systems~\cite{Brambilla1991,Scheuer1999, Desyatnikov2005, Piccardo_NatPhot2022}. Here, we provide evidence that our all-optical lattice supports large scale polaritonic modes (or super-modes), which arise due to the inter-cell interactions in our nonlinear system, rather than to a collection of uncoupled individual single-site states. As such, system states are prone to develop strong instabilities that could demolish their robustness and, hence, their observation. For this reason, the observation of a large number of coupled vortices constitutes in itself a remarkable finding, that could not be anticipated \textit{a priori}.
With 22 cells (Fig.~\ref{fig5}A,B) a large number of stationary states coexist for pump powers just above the condensation threshold and the system exhibits a large degree of multi-stability. As a consequence, the different experimental realizations with 22 cells reveal the formation of a wide diversity of states where individual vortex locations appear stochastically from one realization to the next. 

Future investigations could verify whether it is possible to excite a macroscopic polariton system in a superposition of eigenstates carrying the vorticity and provide a clear answer to a fundamental question: is stochasticity of observed vortex states in such a system a solid proof of its multi-stability? In general, interferometry cannot unambiguously prove that the system is not in a quantum superposition of states within the same energy manifold. Therefore, to verify the hypothesis of a superposition of eigenstates, one would need to implement quantum vortex tomography, which requires experimental single-shot realization of a vortex sorter in free-space and temporal correlation measurements with high resolution. The systematic study of extended lattices of coupled vortices, although challenging, it offers a promising avenue of research with an outlook to quantum simulators that requires thorough theoretical consideration.

In conclusion, we demonstrate evidence of an all-optically imprinted lattice of interacting vortices in nonresonantly driven-dissipative quantum fluids of light. Whereas in a single cell, we observe an intrinsically stochastic behaviour of an optically trapped polariton condensate, in the presence of the second neighbouring cell we reveal synchronization into a vortex-antivortex state, resembling the OAM analogue of AFM coupling. We also analytically predicted and experimentally demonstrated optical control of the vortex interactions from AFM to FM. Therefore, by tuning the interactions between ordered trapped polariton condensates, one can expect observation of various artificial magnetic phases between the nodes, emulating complex magnetic systems, and the study of other exotic coherent states in expanded structures. Our findings address the challenge of realizing the node-to-node coupling in spatially confined quantum fluids of light possessing OAM.  
\\

\section{Materials and Methods}
\textit{Generalised Gross-Pitaevskii model.} The condensate dynamics is assumed to be described by the 2D generalised Gross-Pitaevskii equation (2DGPE) for the polariton wavefunction $\Psi(\textbf{r},t)$ coupled to a rate equation describing the exciton reservoir feeding the condensate $n_{X}(\textbf{r},t)$~\cite{Wouters_PRL2007}:

\begin{eqnarray}
 &&
\nonumber
 		i \hbar \frac{\partial \Psi}{\partial t} = \bigg[-\frac{\hbar^2 \nabla^2}{2m_\mathrm{eff}} - i \frac{\hbar}{2} (\gamma_{c}-Rn_{X}) \\ && + g_{c} |\Psi|^2 + g_{r}(n_{X}+\eta P(\textbf{r})) \bigg] \Psi, \label{eq1} 
 		\\ &&
		\frac{\partial n_{X}}{\partial t} =  -(\gamma_{X}+R|\Psi|^2)n_{X} + P(\textbf{r}). \label{eq2}
\end{eqnarray}
Our parameters are based on the sample properties and past experiments~\cite{Cilibrizzi_APL2014}. Here, $m_\mathrm{eff} \approx 5.63\times10^{-5}m_e$ is the effective mass of the lower-branch polaritons where $m_e$ is the free electron rest mass, $g_{c}=2.4~\upmu \textrm{eV}\upmu\textrm{m}^2$ is the polariton-polariton and $g_{r}=2 g_c$ is the polariton-reservoir interaction strengths typical for GaAs-based systems, $R=0.021~\upmu \textrm{m}^2\textrm{ps}^{-1}$ is fitted rate of stimulated scattering of polaritons from active reservoir, $\gamma_{c}\approx 0.182~\textrm{ps}^{-1}$ and $\gamma_{X}\approx 0.05~\textrm{ps}^{-1}$, are the decay rates for polariton condensate and reservoir excitons, respectively, $\eta \gamma_{X} \approx 0.49$ quantifies an additional blueshift coming from a background of dark reservoir excitons. The function $P(\textbf{r})=(P_0\gamma_{X}\gamma_{c}/R)\sum_{n}Q(\textbf{r}-\textbf{r}_{n})$ describes spatial pump profile consisting of identical Gaussian spots $Q(\textbf{r})=e^{-r^2/d^2}$ of width $d$ (corresponding to $2.5~\upmu\textrm{m}$-wide pump spots used in the experiment) placed in the points with coordinates $\textbf{r}_{n}$ defining particular pump configuration (one or several honeycomb cells or large-scale honeycomb lattice with period $D$). Dimensionless pump amplitude $P_0$ is defined here in units of $\gamma_{X}\gamma_{c}/R$ corresponding to condensation threshold for spatially uniform pump.

The system~\eqref{eq1}-\eqref{eq2} was solved using a variant of split-step fast Fourier transform method combined with fourth-order Runge-Kutta method to account for interaction between the condensate and the reservoir. To uncover all existing stationary states in the system and possible types of evolution, including essentially dynamical regimes, when excitation of stationary states does not occur, the modeling was initiated using different realizations of small-scale noise for initial $\Psi,n_{X}$ distributions. To identify the majority of all possible stationary solutions of the system or dynamical evolution regimes, multiple (up to several hundreds) input noise realizations were implemented for each pump power and pump-configuration. To account for considerable ballistic expansion of polaritons from the pumped regions leading to nonzero polariton density even far from them, we used in modeling sufficiently large spatial domain $300\times300 ~\upmu\textrm{m}^2$ greatly exceeding the pumped area. Stationary states of the system~\eqref{eq1}-\eqref{eq2} corresponding to $\partial_t n_{X} = 0$ were found in the form $\Psi(\textbf{r},t)=w(\textbf{r})e^{-i\varepsilon t}$, where $w(\textbf{r})$ is a complex function describing stationary polariton distribution in the cavity plane, $\varepsilon$ is the energy detuning, using Newton iteration method. Due to the dissipative nature of our system, the energy detuning $\varepsilon$ of stationary states, some of which appear as stable attractors, is also determined by the pump amplitude $P_0$. Stability of such stationary states, whose families are presented in Fig.~\ref{fig4} for different pump configurations, were tested by modeling their evolution in the system~\eqref{eq1}-\eqref{eq2} and by performing linear stability analysis on them. \\

\noindent
\textit{Single condensate vorticity onset.} Here, we will write a variational form of the generalised Gross-Pitaevskii model describing only the dynamics of the first excited degenerate pair of $l = \pm 1$ OAM modes in the condensate. We will then proceed to prove, for an ideal cylindrically symmetric trap, that the only (and equally) stable condensate solutions correspond to these OAM modes.

\begin{figure}
\includegraphics[width=0.97\linewidth]{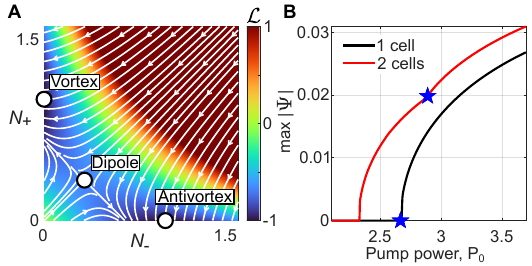}
\caption{\textbf{Vorticity onset in the single and pair of trapped condensate cells.} (\textbf{A}) The colorscale depicts the Lyapunov potential~\eqref{eq.lyp} of the system with two minima (white dots) corresponding to vortex and antivortex solutions in which all phase space trajectories converge towards, underlining the deterministic nature of the system. Here we set $\tilde{p} = \tilde{R} = 1$ without loss of generality. (\textbf{B}) Reproduced results from Fig.~\ref{fig4}A using the variational Gross-Pitaevskii model for a single~\eqref{eq.cpm} and a pair of coupled condensates~\eqref{2osc}. The blue stars denote the onset of vorticity for increasing power.
}
\label{flow}
\end{figure}

We will assume for simplicity, that the cw driven excitonic reservoir follows the condensate adiabatically so that $\partial_t n_X \simeq 0$ which implies,
$n_X = \frac{P(\mathbf{r})}{\gamma_X + R |\Psi|^2} \approx \frac{P(\mathbf{r})}{\gamma_X} \left(1 - \frac{ R |\Psi|^2}{\gamma_X} \right)$.
In the last step we have Taylor expanded the reservoir solution around small $R|\Psi|^2/\gamma_X$ which is valid at pump powers not too far from threshold. This allows us to write a simpler 2DGPE,
\begin{align} \notag
i  \hbar \frac{\partial \Psi}{\partial t} & = \Bigg[ - \frac{\hbar^2 \nabla^2}{2m_\mathrm{eff}} + g_c|\Psi|^2 - \frac{i \hbar \gamma_c}{2}  + g_r \eta P(\mathbf{r}) \\
& + \left(g_r + i \frac{\hbar R}{2}\right) \frac{P(\mathbf{r})}{\gamma_X} \left(1 - \frac{R |\Psi|^2}{\gamma_X} \right) \Bigg] \Psi.
\label{eq.2DGPEv2}
\end{align}
We are interested in a condensate that occupies the degenerate pair of clockwise and anticlockwise OAM states. Assuming an ideal cylindrically symmetric trap $P(\mathbf{r}) = P(r)$ we can write our truncated basis as,
\begin{equation}
\Psi(\mathbf{r}) = \xi(r) \left( \psi_+ e^{i \theta} + \psi_- e^{-i \theta} \right).
\label{eq.basis}
\end{equation}
Here, $\xi(r)$ is the radial steady state ($\partial_t |\Psi|^2 = 0$) part of the condensate in a single optical trap by solving the 2DGPE. Plugging in this truncated solution into~\eqref{eq.2DGPEv2} and integrating over the space, exploiting the orthogonality of the states, we arrive at two coupled equation of motion describing the vortex phase and occupation,
\begin{equation}
i \frac{d \psi_\pm}{d t}  = \left[i \tilde{p} + (\tilde{g}_c - i \tilde{R}) (|\psi_\pm|^2 + 2 |\psi_\mp|^2)   \right] \psi_\pm,
\label{eq.cpm}
\end{equation}
up to an overall energy shift. Note the factor 2 in the counter-rotating nonlinearity (cross-Kerr term) which is orders of magnitude larger than the singlet interaction strength in spinor polariton condensates. The new coefficients are given by,
$\tilde{p}  = \frac{P_0 \gamma_c}{2}  \int \xi(r)^2 Q(r) \, d\mathbf{r} - \frac{ \gamma_c}{2}$, 
$\tilde{g}_c  = \frac{g_c}{\hbar} \int \xi(r)^4 \, d\mathbf{r} -  \frac{P_0 g_r \gamma_c}{\hbar \gamma_X} \int \xi(r)^4 Q(r) \, d\mathbf{r}$, and  $\label{intR}
\tilde{R}  = \frac{P_0 R \gamma_c}{2 \gamma_X}  \int \xi(r)^4 Q(r) \, d\mathbf{r}$.
Equation~\eqref{eq.cpm} can be written in terms of the amplitude and phase of each mode $\psi_{\pm} = \sqrt{N_\pm} e^{i\phi_\pm}$,
\begin{subequations}
\begin{align}
\frac{dN_{\pm}}{dt} &=  2 \left( \tilde{p} - \tilde{R} N_{\pm} - 2\tilde{R} N_{\mp} \right) N_{\pm}, \label{Eq.SingleTrap_a} \\
\frac{d\phi_{\pm}}{dt} &=  \tilde{g}_c( N_{\pm} + 2 N_{\mp}). \label{Eq.SingleTrap_b}  
\end{align} 
\end{subequations}
We see that the change in the phase of the modes is trivially determined by the dynamics of their amplitudes. We therefore only need to focus on solutions of equation~\eqref{Eq.SingleTrap_a}
which has three equilibrium points: $N_\pm=0$ and $N_\mp = \tilde{p}/\tilde{R}$; and $N_+ = N_- = \tilde{p}/3\tilde{R}$. It is easy to show that the only (and equally) stable equilibrium points are the former through the eigenvalues $\lambda$ of the Jacobian of equations~\eqref{Eq.SingleTrap_a}. This can also be visulized by plotting the system Lyapunov potential with extrema corresponding to these solutions,
\begin{equation} 
\mathcal{L} = -2\tilde{p}(N_+ + N_-) -  \tilde{R} (N_+^2 + N_-^2) - 4\tilde{R} N_+ N_-.
\label{eq.lyp}
\end{equation}
The Lyapunov potential satisfies the condition $d \mathcal{L} /dt \leq 0$ which means that all points in phase space flow towards the two minima indicated by the white dots in Fig.~\ref{flow}A corresponding to the two opposite vortex solutions. The unstable saddle point extremum, corresponding to a dipole solution, is also clearly visible. Our simple model evidences that the only stable condensate solutions in a single cell are vortices which is in agreement with experiment.

These vortex solutions bifurcate from the uncondensed normal phase $N_\pm = 0$ at condensation threshold of the single trap  written,
\begin{equation}
    P_{0,\mathrm{thr}}^\mathrm{1 cell} = P_{0,\mathrm{vort}}^\mathrm{1 cell} = \left( \int \xi(r)^2 Q(r) \, d\mathbf{r} \right)^{-1}.
    \label{1cell_vort}
\end{equation}
The above integral quantifies the amount of overlap between the condensate and the pumped region. If this overlap increases, then the threshold is lowered as expected.
\vspace{2mm}

\noindent
\textit{Vorticity onset for two coupled condensates.} The equations describing two coupled trapped condensates read,
\begin{align} \notag
i \frac{d \psi_{n,\pm}}{d t}  & = \left[i \tilde{p} + (\tilde{g}_c - i \tilde{R}) (|\psi_{n,\pm}|^2 + 2 |\psi_{n,\mp}|^2)   \right] \psi_{n,\pm} \\
& + J_a \psi_{3-n,\pm} + J_b \psi_{3-n,\mp},
\label{2osc}
\end{align}
where $n = 1,2$ denote the left and right condensate. Here, $J_{a,b} \in \mathbb{C}$ are the non-Hermitian tunneling rates between co-rotating and counter-rotating vortices (i.e. OAM conserving and non-conserving coupling strengths, respectively),

$J_{a,b}  = \left(\frac{g_r(1+\gamma_X\eta)}{\hbar} + i \frac{R}{2}\right) \frac{P_0 \gamma_c}{R} I_{a,b},$ 

where 
\begin{equation}
    I_{a,b} = \int \xi(r)^* e^{\mp i \theta} Q_\text{tot}(\mathbf{r}) \xi(r') e^{ i \theta'} \, d\mathbf{r}, \quad \in \mathbb{R}.
    \label{intJ2}
\end{equation}
Here, $Q_\text{tot}(\mathbf{r})$ describes the two pumped traps and the separation between them can be written $\mathbf{r}' -\mathbf{r} = d \mathbf{\hat{x}}$. The coordinates of the two pumps are related through $r' = |\mathbf{r}'| = \sqrt{r^2 + d^2 - 2 rd \cos{(\theta)}}$ and $\sin{(\theta')} = r\sin{(\theta)}/r'$. 

Physically, $\text{Re}(J_{a,b})$ and $\text{Im}(J_{a,b})$ correspond to a Josephson (particle conserving) and dissipative (particle non-conserving), respectively, coupling between the condensates. Calculating the integral~\eqref{intJ2} numerically using the single-cell wavefunction found from 2DGPE simulations we find that $I_{a,b}<0$ which means $\text{Im}(J_{a,b})<0$ and therefore the dissipative coupling favours anti-phase synchronisation between neighbouring condensates, see Supplementary Note 7. This is in agreement with observations in experiment Fig.~\ref{fig2}E-H and 2DGPE simulations Fig.~\ref{fig4}E-G. The negative value of the integral intuitively makes sense because anti-phase displaced vortices constructively interfere between the two cells. Thus anti-phase condensates overlap more strongly with the pumped region between the traps and are populated more efficiently.

The lowest threshold solution of the coupled system~\eqref{2osc} is then written $\boldsymbol{\psi} = (\psi_{1,+},\psi_{1,-},\psi_{2,+},\psi_{2,-})^\mathrm{T} = \sqrt{N}(1,1,-1,-1)^\mathrm{T} e^{-i\omega t}$ with occupation $N = [\tilde{p} + |\text{Im}{(J_a + J_b)}|]/3\tilde{R}$ and frequency $\omega = 3N + |\text{Re}{(J_a + J_b)}|$. This is a antiphase dipole-dipole orientated head-to-tails $(\infty) \leftrightarrow (\infty)$ with a condensation threshold of $\tilde{p}_\mathrm{thr} = -|\text{Im}{(J_a + J_b)}|$, in agreement with previous predictions~\cite{Cherotchenko_PRB2021}. Such anisotropic low-threshold solution intuitively makes sense since polaritons flow stronger out of the trap parallel to the condensate dipole axis, thus creating optimal coupling (overlap) conditions between two spatially interacting condensates (e.g., similar to $\sigma$-bonding between atoms). 

At higher pump powers these dipoles in each condensate loose stability and bifurcate each into either $l = \pm 1$ vortices. When $|I_b|>|I_a|$ the system favours AFM ordering, like reported in Fig.~\ref{fig2}E-H. If $|I_b|<|I_a|$ the converse is true and FM ordering is dominant (which we show experimental evidence of in the Supplementary Material). This vortex transition point $P_{0,\mathrm{vort}}^\mathrm{2cells}$ can be analytically derived for the AFM case $|I_b|>|I_a|$ (The FM case follows similar treatment) by locating the onset of linear instability for the dipole-dipole solution,

\begin{widetext}
\begin{equation}
P_{0,\mathrm{vort}}^\mathrm{2cells}  =   \frac{   1  + \dfrac{4 g_c^2(1+\gamma_X \eta)}{R^2}\left(2
- \dfrac{ \gamma_X}{ \gamma_c}  \dfrac{ \int \xi(r)^4 \, d\mathbf{r}}{\int \xi(r)^4 Q(r) \, d\mathbf{r}}     \left(I_a + I_b\right) \right) }{ I_p \left( 1  + \dfrac{8 g_c^2(1+\gamma_X \eta)}{R^2} \right) +  \left( 2I_a + 3I_a \left(\dfrac{4 g_c(1+\gamma_X \eta) }{R} \right)^2  - I_b  - 2  \dfrac{4 g_c^2(1+\gamma_X \eta)}{R^2} \left(I_a + I_b \right) \right) }.
      \label{2cell_vort}
\end{equation}
\end{widetext}
Note that $I_{a,b}<0$ which means that if the coupling anisotropy $I_b/I_a$ is large then the critical power of the AFM vortex formation reduces. This means that by tailoring the pump $P_\mathrm{tot}(\mathbf{r})$ one can adjust $I_{a,b}$ to shift the position of the transition point.

We obtain good results shown in Fig.~\ref{flow}B using $I_{a,b}$ as fitting parameters and solving the variational Gross-Pitaevskii equation for a single cell~\eqref{eq.cpm} and two cells~\eqref{2osc} as a function of increasing power. Our results are in excellent agreement with 2DGPE simulations previously shown in Fig.~\ref{fig4}A, and thus explain the essential observations of our experiment. The blue stars indicate the onset of vorticity in the system. Going further, we have also extended Eq.~\eqref{2osc} to the case of arbitrarily coupled cells and analyzed the onset of vorticity and the preference towards AFM order when $|I_b|>|I_a|$ in the 3-cell and 22-cell configuration (see Supplementary Note 6).

\bibliography{Hexagon}
\vspace{0.5cm}

\noindent
\textbf{Acknowledgements}\\
\textbf{Funding}:
This work was supported by the Russian Science Foundation (RSF) grant no. 21-72-00088. H.S. acknowledges the Icelandic Research Fund (Rannis), grant No. 239552-051. C.M. acknowledges support from the Spanish government via grant PID2021-124618NB-C21 by MCIN/AEI/10.13039/501100011033 and ‘ERDF: a way of making Europe’ of the European Union and the Generalitat Valenciana PROMETEO/2021/082. Y.V.K. acknowledges funding by the research project FFUU-2024-0003 of the Institute of Spectroscopy of the Russian Academy of Sciences.
\\
\noindent
\textbf{Author contributions}:\\
S.A. performed experiments, J.D.T. developed the software for the data acquisition, S.A., K.S. and I.G. analyzed the data, C.M., Y.K., H.S. performed simulations, H.S. developed the variational treatment of the Gross-Pitaevskii model and derived vortex thresholds points, P.L. and Y.K. led the project, all authors contributed to discussions and writing of the manuscript. \\
\noindent
\textbf{Competing interests}:\\
The authors declare that they have no competing interests.\\

\noindent
\textbf{Data and Materials Availability}:\\
All data needed to evaluate the conclusions in the paper are present in the paper and/or the Supplementary Materials.\\

\clearpage

\setcounter{equation}{0}
\setcounter{figure}{0}
\setcounter{table}{0}
\setcounter{page}{1}


\renewcommand{\theequation}{S\arabic{equation}}
\renewcommand{\thefigure}{S\arabic{figure}}
\renewcommand{\thesection}{S\arabic{section}}

\onecolumngrid

\newpage
\begin{center}
\Large\textbf{Supplementary Materials}
\end{center}

\section*{Supplementary Note 1: Lattice constant dependence}
\begin{figure}[h!]
    \includegraphics{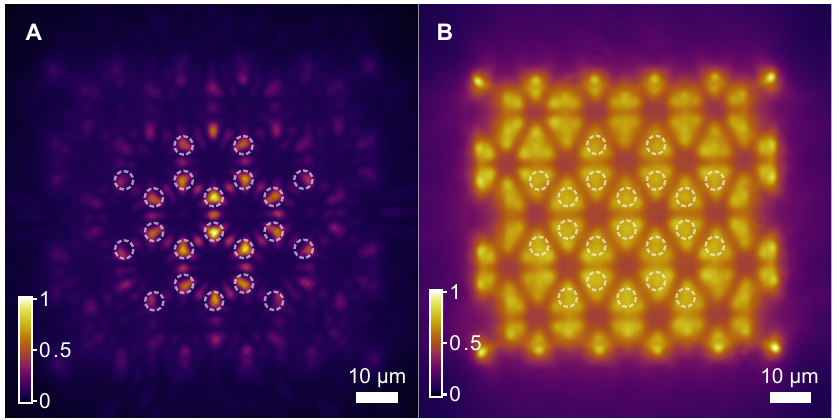}
    \caption{\textbf{Lattice constant dependence.} (\textbf{A}),(\textbf{B}) Experimentally measured time-averaged polariton photoluminescence intensity distribution for the lattice pumped above condensation threshold, and lattice constant set to $D=15.3~\upmu$m and $D=14.1 ~\upmu$m, respectively. In (\textbf{A}) the condensates form on top of the pumping spots and synchronize across the lattice, while for smaller lattice constant in (\textbf{B}) the polaritons repelled outside pumped areas form triangular-shaped condensates. White dashed circles schematically denote pumped spots, forming honeycomb lattice.}
    \label{fig.S1}
\end{figure}

\vspace{0.20 cm}
In Fig.~\ref{fig.S1} we show the real-space polariton photoluminescence (PL) excited nonresonantly in fragment of honeycomb lattice. When the lattice constant is set to $D=$ 15.3 $\upmu$m, we observe formation of ballistically coupled polariton condensates, co-localised with Gaussian pumps as visible from Fig.~\ref{fig.S1}A. For smaller lattice constant set to $D=14.1 ~\upmu$m we clearly observe repulsion of polaritons outside the pumped spots, which leads to formation of triangular-shaped (Fig.~\ref{fig.S1}B) polariton condensates above threshold. Finally, when $D=$ 11.5 $\upmu$m, our time-integrated measurements reveal formation of large-scale vortex lattice, shown in Fig.~5A of the manuscript.

\newpage
\section*{Supplementary Note2: Pump profile used in experiments with ``building blocks'' of the lattice}
\begin{figure}[h!]
    \includegraphics[scale=0.93]{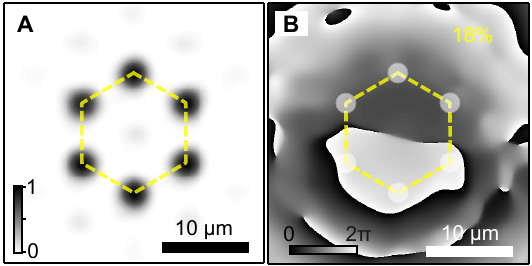}
    \caption{\textbf{Nonresonant excitation profile for a single cell and an example of a dipole state.} (\textbf{A}) Measured laser pattern with balanced intensities of the spots and (\textbf{B}) measured single-shot phase map, corresponding to a dipole state (observed in 18 realizations out of 100). Dashed lines and semi-transparent white circles schematically denote pump spots.}
    \label{fig.S2}
\end{figure}

\vspace{0.2 cm}
In order to realize a truly spontaneous formation of the vortex states in one-, two- and three-cell structures, we use active feedback technique that allows for equalizing the integrated intensities of the pump spots with a typical achievable precision (standard deviation) below 2\%. For this we real-time measure a corresponding pump profile, extract pumps intensities  and iteratively recalculate a phase mask, applied to spatial light modulator (SLM), to reach desired uniformity of the pump pattern. When succeeded, the calculation stops and the SLM phase mask becomes static (i.e. no longer updated). At this stage all experiments for specific structure are conducted: time-integrated PL measurements and single-shot homodyne interferometry. This approach ensures that the excitation laser profile remains exactly the same throughout measurements. As an example in Fig.~\ref{fig.S2}A we show obtained pump profile used in experiments with a single honeycomb cell. The statistical occurrence for vortex and antivortex states, shown in Fig.~1D and Fig.~1E, convincingly proves that pump profile is uniform to avoid pump-induced vorticity with preferable topological charge. However, out of 100 single-shot realizations we observed also 18 realizations of the dipole states (see Fig.~\ref{fig.S2}B), which could appear due to sample disorder and/or finite imperfections of pump profile (though balanced in intensity). The numbers, in yellow, shown in Fig.~1D and Fig.~1E of the manuscript are renormalized such as to show statistical occurrence out of 82 realizations.  

To confirm that selected number of one hundred shots is reasonable enough for statistics to provide a trustful distribution of cases occurrence, we analyzed 401 single-shot realizations for 2-cell system shown in Fig.~2 of manuscript and obtained very similar occurrence (90.8\% vs 93\% obtained for 100 shots) for realizations with opposite OAM. The rest 9.2\% correspond to observation of vortex-dipole (or antivortex-dipole) states. 
\begin{figure}
    \includegraphics[scale=0.9]{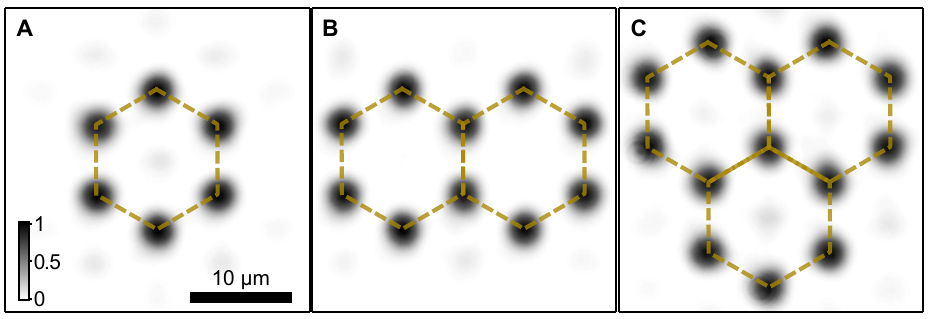}
    \caption{\textbf{Nonresonant excitation profile for ``building blocks'' of the lattice with constant set $D=$ 11.5 $\upmu$m.} Normalized experimentally measured laser profile used to excite vortices in one (\textbf{A}), two (\textbf{B}) and three (\textbf{C}) honeycomb cells, corresponding to experimental data shown in Fig.~1C,D,E, Fig.~2 and Fig.~3 of the manuscript. Dashed lines schematically denote the excitation pattern to guide the eye. Scale bar in (\textbf{A}) applies to all panels.}
    \label{fig.S3}
\end{figure}
All nonresonant pump profiles used in the experiments with one-, two- and three-cell structures are shown in Fig.~\ref{fig.S3}. The characteristic size of the unit cells is kept the same ($D=$ 11.5 $\upmu$m).

\newpage
\section*{Supplementary Note 3: Homodyne interferometry technique}
\vspace{0.2 cm}
In this work we use nonresonant continuous wave (CW) excitation laser, spatially shaped according to desired excitation pattern (number of nodes, their diameter, separation distance and relative intensities). In order to reconstruct a real-space polariton PL intensity and the corresponding phase map of the condensate state from measured single-shot interference pattern (obtained with a homodyne interferometry), we apply the off-axis digital holography technique. 

\begin{figure}[h!]
    \includegraphics{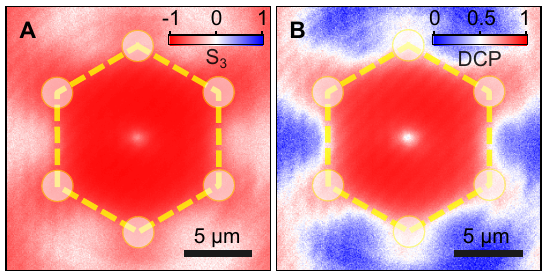}
    \caption{\textbf{Polarization-resolved polariton photoluminescence in a single cell.} (\textbf{A}) Experimentally measured map of the Stokes parameter $S_{3}$ for the trapped condensate excited with CW circularly polarized laser emission. Corresponding degree of circular polarization (DCP) map in \textbf{B} confirms that the condensate is mostly circularly polarized. Dashed lines with circles schematically denote the excitation pattern.}
    \label{fig.L2}
\end{figure}

\begin{figure}[h!]
    \includegraphics{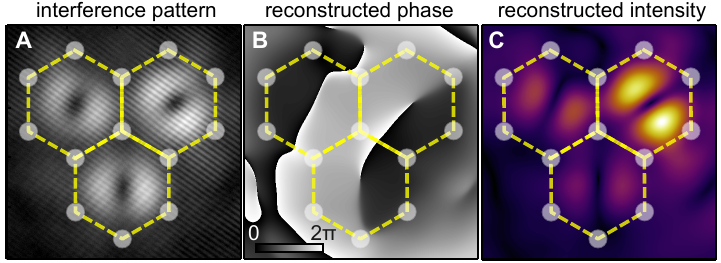}
    \caption{\textbf{Analysis of example single-shot interference pattern for 3-cell configuration.} (\textbf{A}) Recorded interference pattern, (\textbf{B}) corresponding reconstructed phase map and (\textbf{C}) reconstructed intensity map. Semi-transparent white circles denote pump positions to guide the eye.}
    \label{fig.S4}
\end{figure}

Below we briefly describe the principle of homodyne interferometry method. First, in addition to acousto-optically modulated CW laser, we use a weak CW seed resonant laser, synchronously modulated to the nonresonant pump laser. In all our experiments, described in the manuscript, we used circularly polarized laser for optical excitation.  As we show in Fig.~\ref{fig.L2}, at pump power used in the experiments, a condensate trapped inside a hexagonal cell inherits the polarization from the pump even under nonresonant excitation. This allows us to excite a single energy state corresponding to polaritons with a single predominant pseudo-spin and make therefore, a ``clean'' system without any energy splitting (between $\sigma^+$ and $\sigma^-$ polaritons).
The seed laser, tuned to the energy of the condensate, is focused (FWHM $\approx$ 2 $\upmu$m) on the sample through the same microscope objective as the main excitation pattern, and essential only to locally fix the phase of the condensate with respect to the reference wave. Second, the collected polariton PL, is interfered with a flat reference wave in Mach-Zehnder interferometer. The reference wave originates from the same laser (with narrow linewidth of $\approx 100$ kHz) as the seed. Third, the interference pattern (see Fig.~\ref{fig.S4}A) is recorded with a sensitive CCD camera and analyzed using off-axis digital holography. The resultant real-space phase and intensity maps are shown in Fig.~\ref{fig.S4}B,C.  

\begin{figure}
    \includegraphics[scale=0.9]{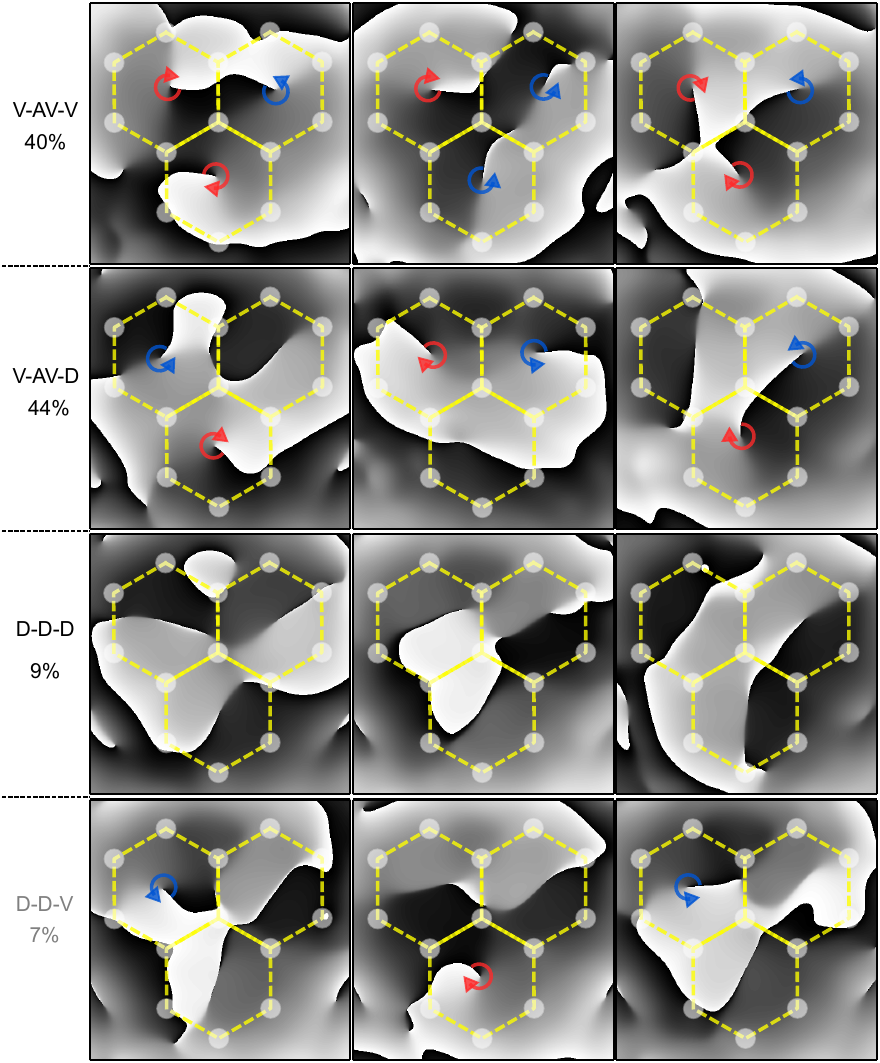}
    \caption{\textbf{Examples of the reconstructed phase maps corresponding to different families of the states observed in 3-cell structure.} The number in \% shows corresponding statistical occurrence based on analysis of one-hundred single shot interference patterns. Semi-transparent white circles denote pump positions to guide the eye. V, AV and D stands for vortex, antivortex and dipole states, respectively.}
    \label{fig.S5}
\end{figure}

Figure~\ref{fig.S5} shows different examples of experimentally measured real-space phase maps corresponding to predicted families of the states observed in 3-cell structure (shown in Fig.~3 of the manuscript). The numbers here correspond to statistical occurrence of the states over one hundred individual realizations. We note that only in 7\% of cases (7 realizations) we classified the state as dipole-dipole-vortex (bottom panels in Fig.~\ref{fig.S5}). We believe this state can be explained by some tiny imperfections of the pump or sample disorder given the nonlinear nature of exciton-polariton system. The rest of the families are in a good agreement in our theoretical prediction. The numbers, in yellow, shown in Fig.~3B-D of the manuscript are renormalized such as to show statistical occurrence out of 93 single-shot realizations. 

\newpage
\section*{Supplementary Note 4: Interaction control in coupled cells}
\begin{figure}[h!]
    \includegraphics[scale=0.86]{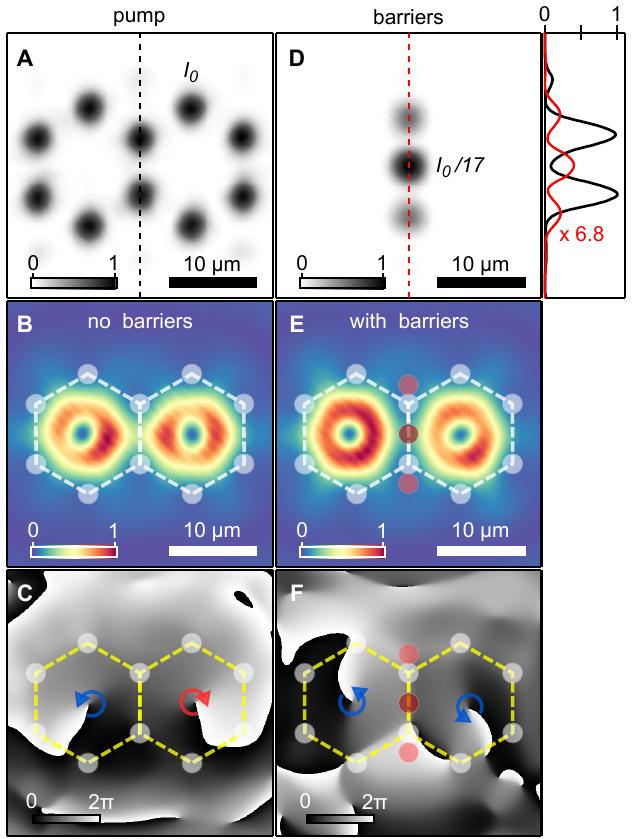}
    \caption{\textbf{Switching the vortex interactions by means of weak optically imprinted barriers in between the cells.} (\textbf{A}) Normalized nonresonant pump profile and (\textbf{B}) corresponding time-averaged polariton PL, demonstrating formation of (\textbf{C}) vortex-antivortex pair. When the barriers pump in (\textbf{D}) additionally injected, the system switches to vortex-vortex state as visible from PL intensity in (\textbf{E}), and single-shot pahse map in (\textbf{F}). White circles schematically denote the condensates pump positions, red circle denote barriers positions.}
    \label{fig.S6}
\end{figure}

\vspace{0.2 cm}
Our experimental measurements of the single-shot realizations in 2-cell structure revealed that such system preferentially occupies vortex-antivortex or antivortex-vortex states. However, none of the realizations demonstrated the same orbital angular momentum (OAM). At the same time, our developed model predicts that by changing the coupling between the cells one can potentially tune the system to vortex-vortex state, implying FM order instead of dominant AFM arrangement. To realize this experimentally, we first reproduce already expected vortex-antivortex pair using pump profile shown in Fig.~\ref{fig.S6}A. As a result, above condensation threshold power we observe formation donut-shaped PL inside the cells (Fig.~\ref{fig.S6}B), corresponding to a vortex-antivortex pair as confirmed by the extracted phase map in Fig.~\ref{fig.S6}C. 

To change the coupling between the trapped synchronized states we additionally optically inject three nonresonant weak excitation beams, as shown in Fig.~\ref{fig.S6}D (see also inset on the right with the line profiles of the condensate pumps and barriers pumps). We note, that for the barrier pattern we use cross-circularly polarized excitation with respect to the condensate pumps. This allows to minimize gain due to overlap of the condensate wave functions with the barrier potentials in the middle of the 2-cell structure. It turns out, that even when the brightest spot of the barrier pattern (Fig.~\ref{fig.S6}D) is $\approx$17 times less intense ($I_0$/17) than average integrated intensity $I_0$ of the condensate pump (Fig.~\ref{fig.S6}A), the initial system state definitely switches to vortex-vortex, as shown in Fig.~\ref{fig.S6}E,F. By this we mean that in none of the analyzed 100 single-shot realizations we observed states with opposite OAM of polaritons in the traps, once the barrier pump overlaid with condensate pump. Such an all-optical approach with structured excitation to control the vortices interactions opens a new route towards engineering the macroscopic coherent states with desired spatial distribution of OAM. 

\newpage
\section*{Supplementary Note 5: Experimental evidence of ``orbital antiferromagnetism'' ordering in 22-cell structure}
\begin{figure}[h!]
    \includegraphics[width=0.97\linewidth]{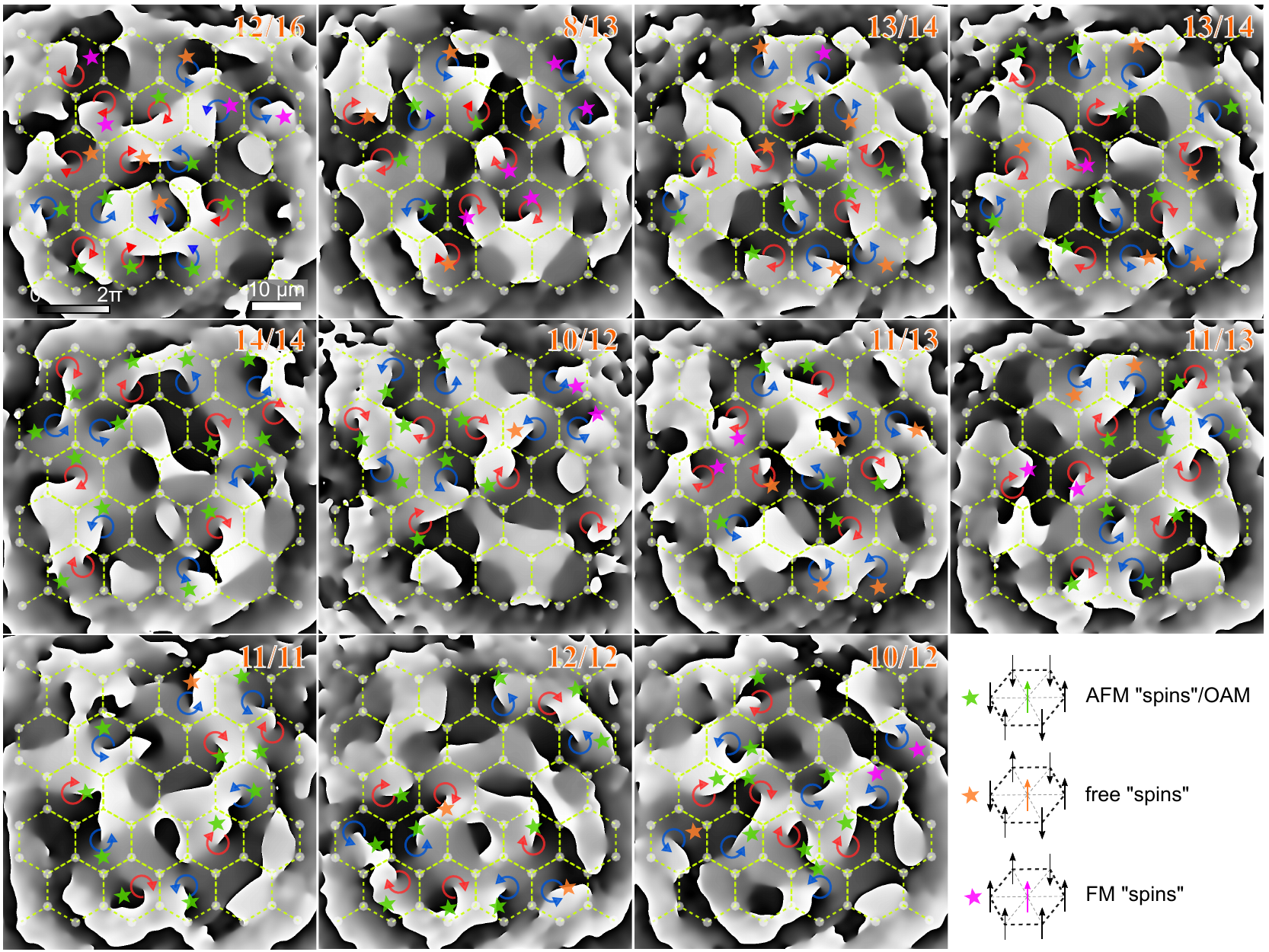}
    \caption{\textbf{Experimentally extracted single-shot phase maps of the polariton condensate in 22-cell structure with lattice constant set $D=11.5~\upmu$m.} Blue and red arrows schematically indicate vortices and antivortices. Each lattice cell with vortex/antivortex inside is classified according to the description given in the main manuscript and marked by a star. Green star corresponds to ``spin''(vortex) which has AFM coupling with neighbours. Orange star corresponds to free ``spin'', or total zero OAM of polaritons in neighboring cells. Magenta star corresponds to ``spins'' with dominantly co-aligned neighbours (FM spin). Dashed lines with semi-transparent white circles denote pump pattern to guide the eye.}
    \label{fig.S7}
\end{figure}

\vspace{0.2 cm}
Figure~\ref{fig.S7} shows examples of extracted from experiment single-shot phase maps corresponding to the state of 22-cell structure, shown in Fig.~5A of the manuscript. With dashed lines and semi-transparent circles we schematically denote nonresonant pump pattern. Each shown phase map reveals numerous vortices (follow blue arrows) and antivortices (follow red arrows). The total number of phase dislocations associated with vortices and antivortices, localized inside the honeycomb cells, is written in the right top corner, in orange, as the second number. The first written number, in orange, shows the sum of classified AFM and free nodes/vortices (by analogy with spins) as described in the main manuscript. For visual perception, in each cell with vortex or antivortex we place a star with a colour, corresponding to spin category (see small insets at the bottom in Fig.~\ref{fig.S7}). The analysis clearly indicates that dominant portion of cells with vortices demonstrate AFM coupling (i.e. major part on neighbours has opposite measured OAM of polaritons). Totally, we analyzed 25 single-shot realizations with the results plotted in Fig.~5C as empty markers. Remarkably, such approach allows to experimentally evidence the presence of dominant AFM order across different realizations with lacking any order at first glance.

\begin{figure}
    \includegraphics{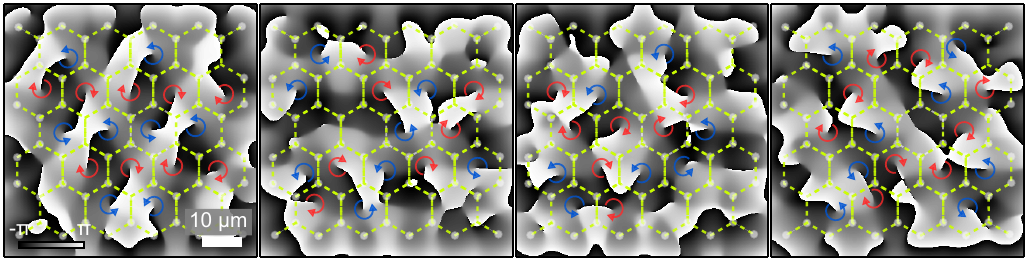}
    \caption{\textbf{Simulated single-shot realizations of the condensate phase map for the 22-cell structure.} Dashed lines with semi-transparent white circles schematically denote pump pattern.}
    \label{fig.S8}
\end{figure}

\newpage
Figure~\ref{fig.S8} shows simulated instantaneous phase maps, corresponding to Fig.~5D of the manuscript. Similar to experimental observations in Fig.~\ref{fig.S7} the vortices stochastically flip their topological charges from realization-to-realization, yet maintaining the dominant AFM order across 22-cell structure. In simple words, the polaritons OAM in the given cell strongly depends on the neighbouring cells.    

\section*{Supplementary Note 6: Variational Gross-Pitaevskii model for coupled polariton vortices}
\vspace{0.2 cm}
Extending Eq.~(10) in the main text to arbitrary geometries of coupled nearest-neighbour polariton vortices gives,
\begin{equation}
i \frac{d\psi_{n,\pm}}{d t}  = \left[i \tilde{p} + (\tilde{\alpha} - i\tilde{R} ) (|\psi_{n,\pm}|^2 + 2 |\psi_{n,\mp}|^2)   \right] \psi_{n,\pm} + \sum_{\langle n,m \rangle} \left[J_a \psi_{m,\pm} + J_b \psi_{m,\mp} e^{\mp 2i \Theta_{n,m}}\right].
\label{eq.latt}
\end{equation}
Here, $\psi_{n,\pm}$ is the phase and amplitude of the $n$th condensate component with OAM $l=\pm 1$, $J_{a,b} \in \mathbb{C}$ are the tunneling rates between co-rotating and counter-rotating vortices, $\tilde{\alpha}$ corresponds to the repulsive polariton-polariton interactions, $\tilde{R}$ represents a gain saturation mechanism in the adiabatic exciton-reservoir limit, and $\tilde{p}$ is the combined non-resonant optical pumping rate and cavity losses. The sum runs over nearest neighbours and $\Theta_{n,m}$ is the angle of the link between two condensates in separate traps [notice the double winding in the exponent of Eq.~\eqref{eq.latt}]~[47]. In the special case of triangular geometry we have $\Theta_{n,m} \in \{0, 2\pi/3, 4\pi/3\}$.

\subsection{3-cell system: AFM order in a single triangle}
\vspace{0.2 cm}
Here we numerically investigate the presence of AFM order in a single triangle when the power parameter $\tilde{p}$ is scanned. It is convenient to characterise the behaviour of the three-cell system by defining a three-dimensional Bloch vector (or pseudospin) for each condensate similar to what is done in optics with light beams carrying OAM,
\begin{equation}
    S_{x,y,z}^n = \begin{pmatrix} \psi_{n,+}^* & \psi_{n,-}^* \end{pmatrix} \hat{\sigma}_{x,y,z} \begin{pmatrix} \psi_{n,+} \\ \psi_{n,-} \end{pmatrix},
\end{equation}
where $\hat{\sigma}_{x,y,z}$ are the three Pauli matrices. Physically, projection on the $S_{x,y}^n$ components means that the $n$th condensate has some dipolar structure. Projection on the $S_z^n$ component means that the $n$th condensate has vorticity. Specifically, $S_z^n>0$ corresponds to a counterclockwise vortex and $S_z^n<0$ to a clockwise vortex.

\begin{figure}
    \centering
    \includegraphics[width=0.98\linewidth]{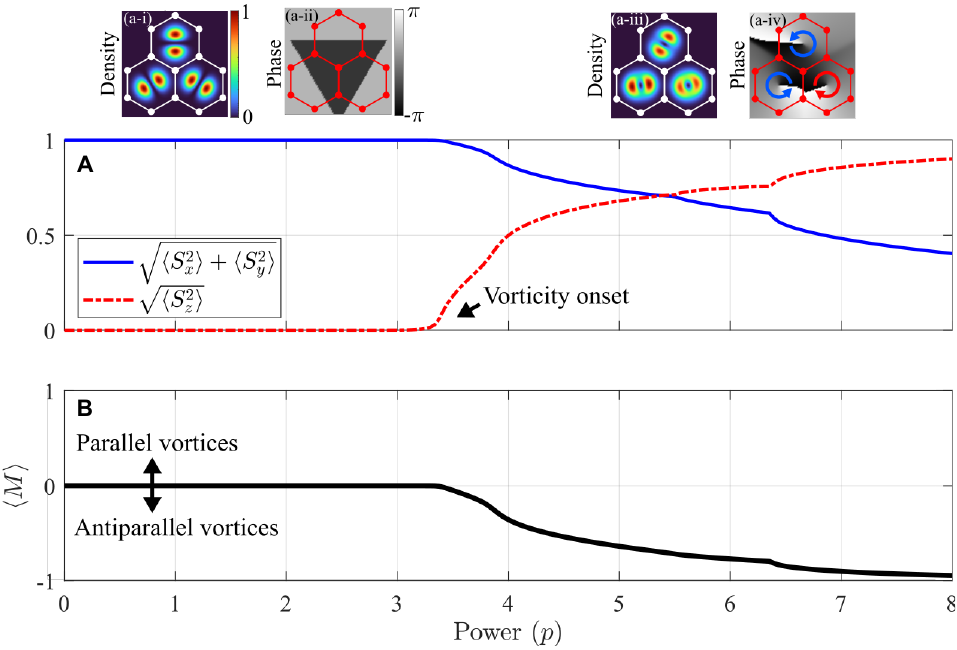}
    \caption{\textbf{Comparison of the average condensate population in dipole states ($S_{x,y}$) and vortex states ($S_z$)}. We numerically solve Eq.~\eqref{eq.latt} over 1000 random initial conditions for different values of the effective power parameter $\tilde{p}$. \textbf{a} Shows the time-average and cell-average magnitude of projection onto the $S_{x,y}$ and $S_z$ Bloch vector components. Left and right insets above (\textbf{A}) show examples states from simulation for low and high power respectively. (\textbf{B}) Corresponding average vortex “magnetism” order parameter. Positive values indicate vortex “ferromagnetism” and negative values vortex “antiferromagnetism”.}
    \label{fignote6}
\end{figure}

We numerically solve Eq.~\eqref{eq.latt} for the triangle of traps, averaging over 1000 random initial conditions. We show in Fig.~\ref{fignote6}A the normalised time-average and cell-average projection of the condensates on the dipole states $\sqrt{\langle S_x^2 \rangle + \langle S_y^2 \rangle}$ and vortex states $\sqrt{\langle S_z^2 \rangle}$ where,
\begin{equation}
    \langle S_{x,y,z}^2 \rangle = \frac{1}{3} \sum_{n=1}^3 \frac{1}{T} \int_0^T \left(S_{x,y,z}^n\right)^2 \, dt,
\end{equation}
as a function of pump power $\tilde{p}$ scaled in units of $|\text{Re}{(J_a)}|$. Here we set $\text{Re}{(J_{a,b})}<0$ and $\text{Im}{(J_{a,b})}<0$ and $|J_b|/|J_a| = 1.5$ in agreement with the overlap integral calculation shown in Fig.~\ref{fig.coupl}. We choose $\text{Im}{(J_{a,b})}/\text{Re}{(J_{a,b})} = 0.5$, and $\tilde{\alpha} = 0.1 \tilde{R}$. We point out that our results do not depend strongly on the choice of parameters as long as $\text{Im}{(J_a)}<\text{Im}{(J_b)}<0$.

From an optics perspective, these quantities are also similar to a light source’s degree of linear polarization (DLP) and degree of circular polarization (DCP) except now for OAM. We also show in Fig.~\ref{fignote6}B the “amount” of AFM order using the following order parameter,
\begin{equation}
    \langle M \rangle = \sum_{n<m} \frac{1}{T} \int_0^T S_z^n S_z^m \, dt.
\end{equation}
Namely, if the order parameter $\langle M \rangle <0$ then the system is preferentially AFM aligned. This quantity is useful when the condensate dynamics are nonstationary, and time-averaging over a long time window $T$ is more meaningful.

The results in Fig.~\ref{fignote6} show two distinct regions of interest. At low powers a single fixed point solution is dominant corresponding to dipole condensates arranged 120$^\circ$ with respect to each other (see insets a-i and a-ii of example solution projected into the spatial domain for clarity), reminiscent of an XY ground state. In this regime there is no vorticity and therefore $\langle M \rangle = 0$. At higher powers we pass the vorticity threshold and vorticity starts growing monotonically. Here, a new family of attractors in which (on-average) two parallel vortices and one antiparallel appear (see insets a-iii and a-iv for example solution). For the given parameters $|J_a|<|J_b|$ the vorticity threshold is associated with clear AFM order as can be seen in Fig.~\ref{fignote6}B. These results are similar to the observations on coupled nanodisk lasers (see Fig.~2 in Ref.~[40]) Notice also that this solution lacks discrete rotational- or mirror-symmetry just like we observe in experiment (see e.g. Fig.~3D in main text).

\subsection{22-cell system: AFM order in a triangular lattice}
\vspace{0.2 cm}
We next investigate for AFM order in the finite 22-cell triangular lattice like in Fig.~5 in the main manuscript by solving Eq.~\eqref{eq.latt}. We use the same parameters as in previous section but now fix $\tilde{p}=1$ which is well above threshold and the vortex bifuraction point in the lattice. The reason why $\tilde{p}=1$ is sufficient is because of the increased connectivity in the lattice compared to the single triangle (i.e., 6 neighbours in the bulk instead of just 2) which lowers the bifurcation point.

We then repeat the same analysis as we did in the last part of the main manuscript where we look for correlations between the obtained vortex patterns and the configurations of the Ising Hamiltonian [Eq.~(1) in the main manuscript]. To do this, we numerically integrated Eq.~\eqref{eq.latt} in time (over a long enough time interval to capture long-time behaviour) and extracted the average $S_z$ parameter from each cell. 
\begin{equation}
    \langle S_z^n \rangle = \frac{1}{T} \int_0^T S_z^n \, dt,
\end{equation}
We then assigned a binary variable to each cell through the following projection,
\begin{equation}
    \sigma_n = \text{sign}{\langle S_z^n \rangle }.
\end{equation}
Repeating this for 1000 random different initial conditions we performed the same analysis of finding the average number of AFM, FM and FREE ``spins'' and added the result to Fig.~5C in the main text.

\section*{Supplementary Note 7: Interference between vortices and optimal phase-relation}
\begin{figure}[t!]
    \centering
    \includegraphics[width=0.6\linewidth]{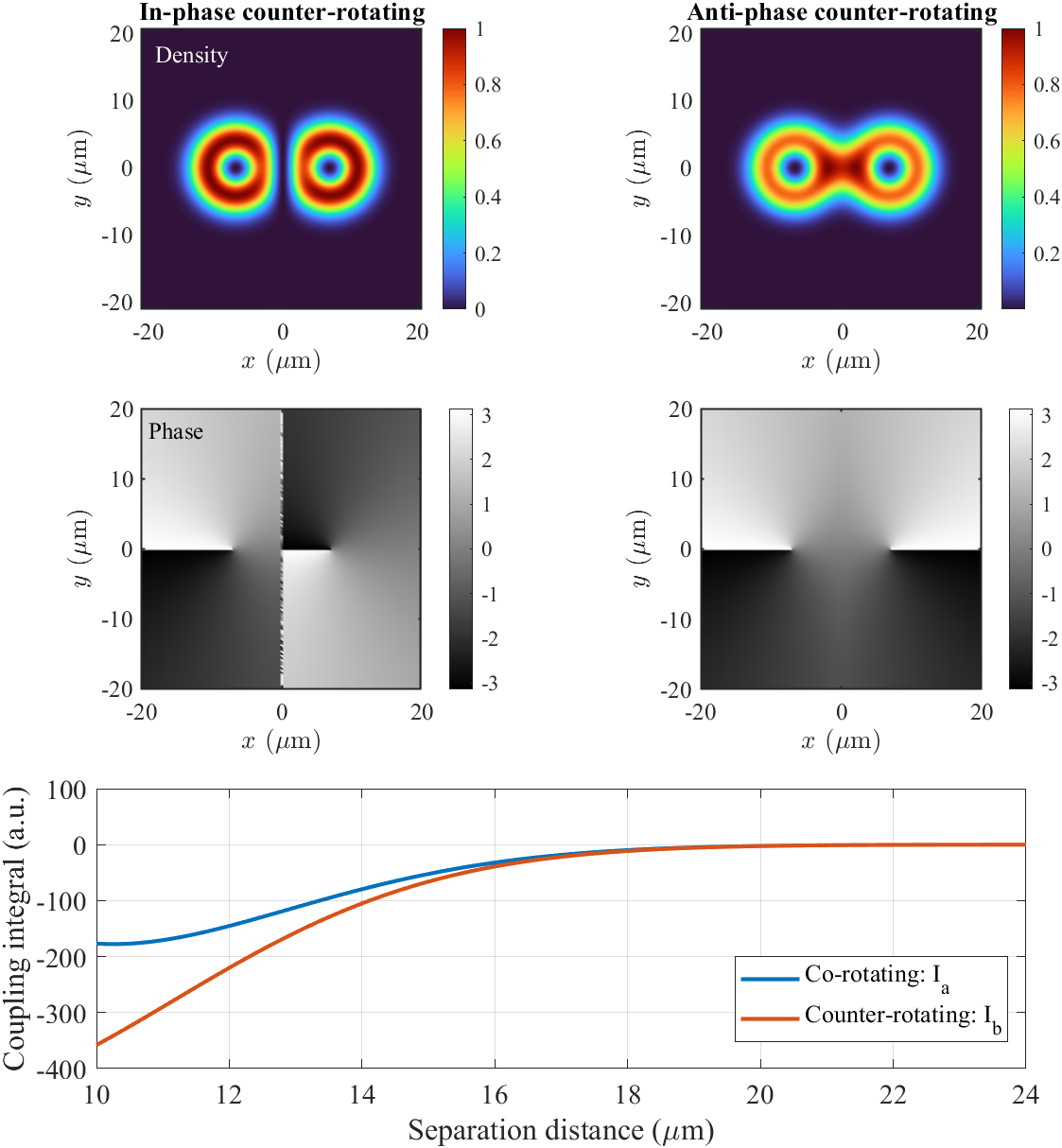}
    \caption{Example superposition of counter-rotating spatially displaced vortex wavefunctions with in-phase (left upper panels) and anti-phase (right upper panels) configuration. Bottom panel shows the coupling integral corresponding to Eq.~(11) in the main manuscript.}
    \label{fig.coupl}
\end{figure}

\vspace{0.2 cm}
The relative phase between the vortices is an important degree of freedom as we discuss around the overlap integral, Eq.(11), in the main manuscript. Here, we numerically calculate this integral between two displaced vortices,
\begin{equation}
    I_{a,b} = \int \xi(r)^* e^{\mp i \theta} \xi(r') e^{ i \theta'} \, d\mathbf{r}.
    \label{intJ2}
\end{equation}
Here, $(r,\theta)$ are the radial and angular coordinate in the cavity plane and $\xi(r)$ are the profiles of each vortex separated by a distance $\mathbf{r}' -\mathbf{r} = d \mathbf{\hat{x}}$. The coordinates of the two pumps are related through $r' = |\mathbf{r}'| = \sqrt{r^2 + d^2 - 2 rd \cos{(\theta)}}$ and $\sin{(\theta')} = r\sin{(\theta)}/r'$. Example profiles for counter-rotating vortices are shown in the top panels in Fig.~\ref{fig.coupl}. Clearly, the in-phase vortices destructively interfere around the mirror symmetry axis where their overlap over the pumped region is most important. In contrast, the anti-phase vortices constructively interfere which optimizes their mutual gain. This can also be seen from calculating their overlap integral as a function of separation distance (lowest panel) where the larger negative value of the counter-rotating vortices (red curve) implies preference towards anti-phase locking in order to optimize the system gain. This explains why the latter configuration has lowest condensation threshold according to analysis of the eigenmodes of 2 cells and why it is observed in the majority of realizations of such pump landscape in the experiment.

\section*{Supplementary Note 8: Gain optimization of the vortex lattice and Ising order}
\vspace{0.2 cm}
In what follows we provide an argument for the Ising analysis in the main manuscript from a theoretical point of view. In order to do this we must construct a cost function for the gain between coupled condensates. We will do this based on Eq.~\eqref{eq.latt} which describes each vortex as a two-component phase-amplitude oscillator in a discrete Gross-Pitaevskii model. The complex-number order parameter for the $n$th condensate is written $\boldsymbol{\psi}_n \equiv (\psi_{n,+},\psi_{n,-})^T$, where $\pm$ denote the right-hand and left-hand rotating vortices. In the work (\textit{37}) the phases $\theta_n$ of coupled scalar condensates are regarded as a 2D classical XY ``spins'' $\mathbf{s}_n = (\cos{(\theta_n)}, \sin{(\theta_n)})$ in the complex plane. Since, in our work, we have a two-component complex state vector $\boldsymbol{\psi}_n \in \mathbb{C}^2$ we need to keep track of both intra- and inter-phases using instead a 4D classical ``spin'', $\mathbf{s}_n = (a_{n,+},b_{n,+},a_{n,-},b_{n,-})^T$ where $a_{n,\pm}$ and $b_{n,\pm}$ are the real and imaginary parts of $\psi_{n,\pm}$. A straightforward calculation gives us the following energy functional corresponding to the coupling energy between condensates [last term in Eq.~\eqref{eq.latt}],
\begin{equation} \label{eq.cost}
    \mathcal{J} = \sum_{n,m} J_{n,m}^{\alpha,\beta} \left[(\mathbf{s}_n \cdot \mathbf{s}_m) \delta_{\alpha,\beta} + (\mathbf{s}_n \cdot \left[ (\hat{\sigma}_1 \otimes \hat{\sigma}_0) \cos{(2\Theta_{n,m})} - (\hat{\sigma}_2 \otimes \hat{\sigma}_2) \sin{(2\Theta_{n,m})} \right] \mathbf{s}_m ) \delta_{\alpha,-\beta}  \right]
\end{equation}
Here, $\alpha,\beta \in \{\pm\}$ are the vortex sign indices, the $\delta_{\alpha,\beta}$ is the Kronecker delta function, $\hat{\sigma}_{1,2}$ are the Pauli matrices and $\hat{\sigma}_0$ is the $2\times2$ identity matrix. The coupling coefficient $J_{n,m}^{\alpha,\beta} \in \mathbb{C}$ is determined by the integrals in Eq.~(11) in the main manuscript, and $\Theta_{n,m}$ is the angle of the link between two condensates in the plane. Notice that in the case of a triangular lattice the angles belong the set $\Theta_{n,m} \in \{ 0, 2\pi/3, 4\pi/3\}$ and are responsible for the geometric frustration. 

As we have discussed in the Materials and Methods section of the main manuscript around Eq.~(11), the coupling rate between condensates $J_{n,m}^{\alpha,\beta}$ is complex which means that $\mathcal{J}$ is complex. Since the system selects an OAM configuration to optimize the gain then we are interested in maximizing $\text{Im}{(\mathcal{J})}$. From here on we are only concerned with the complex part of $\mathcal{J}$ and will drop the ``Im(.)'' notation.

The first inner product term in Eq.~\eqref{eq.cost} favours co-rotating vortices (FM arrangement). However, this term turns out to be weaker than the second term (see  Fig.~\ref{fig.coupl}) which instead favours AFM arrangement. For simplicity, we will investigate the case of $\Theta_{n,m}=0$ corresponding to a non-frustrated linear chain of condensates with uniform coupling strengths $J_{n,m}^{+,-} = J_b$ and $J_{n,m}^{+,+} = J_a = 0$. We then have,
\begin{equation} \label{eq.cost2}
    \mathcal{J} =  J_b \sum_{n,m}  \mathbf{s}_n \cdot (\hat{\sigma}_1 \otimes \hat{\sigma}_0)   \mathbf{s}_m. 
\end{equation}
When pumped strongly enough, the presence of dipole states diminishes (i.e., above blue stars in Fig.~4 in the main manuscript) and we can parametrize the state vector as follows,
\begin{equation}
\mathbf{s}_n = (\cos{(\theta_n)} \cos{(\phi_n)}, \cos{(\theta_n)} \sin{(\phi_n)},  \sin{(\theta_n)} \cos{(\phi_n)},\sin{(\theta_n)} \sin{(\phi_n)})^\text{T}
\end{equation}
Here, $\theta_n \in [0,\pi/2]$ determines the direction and amount of vorticity and $\phi_n \in [0,2\pi)$ the overall phase at site $n$. That is, $\theta_n=0$ corresponds to OAM$=+1$ and $\theta_n = \pi/2$ to OAM$=-1$. Our gain functional then becomes,
\begin{equation} \label{eq.cost3}
    \mathcal{J} =  J_b \sum_{n,m}  \sin{(\theta_m + \theta_n)} \cos{(\phi_m - \phi_n)}. 
\end{equation}
The cosine term is the XY-Hamiltonian previously pointed out in the work (\textit{37}). The sine term is novel and importantly only depends on the angle $\theta_n$ which determines the vorticity in each component. 

If we project the vortex angles onto their nearest extremes $\theta_n \to \theta_n \in \{0, \pi/2\}$ we can rewrite our gain functional using a binary variable $\sigma_n \in \{ \pm \}$ denoting the projected normalized OAM at each site,
\begin{equation}\label{eq.cost4}
 \mathcal{J} = J_b \sum_{n,m} \frac{1 - \sigma_n \sigma_m}{2} \cos{(\phi_m - \phi_n)}.    
\end{equation}
We remind that $J_b<0$ and optimizing~\eqref{eq.cost4}, i.e. optimizing the system gain, is the same as minimizing Eq.~(1) in the main manuscript because they differ by a sign factor. Notice that the binarized term $1 - \sigma_n \sigma_m>0$ in the sum is always positive. Therefore, the optimal value is obtained when the ``Ising'' term is as positive as possible and the cosine term is as negative as possible. This happens for anti-phase $\theta_n-\theta_m=\pi$ and AFM ordered vortices $\sigma_n \sigma_m=-1$, in agreement with our experimental and numerical results.

\end{document}